\documentclass[%
 reprint,
superscriptaddress,
nofootinbib,
nobibnotes,
 amsmath,amssymb,
 aps,
prb,
longbibliography
]{revtex4-2}

\usepackage{graphicx}% Include figure files
\usepackage{dcolumn}% Align table columns on decimal point
\usepackage{bm}% bold math
\usepackage[colorlinks=true,citecolor=blue,linkcolor=blue, urlcolor=cyan]{hyperref} 
\usepackage{empheq}
\usepackage{xcolor}
\usepackage[normalem]{ulem}
\usepackage[english]{babel}

\usepackage[T2A]{fontenc}
\usepackage[utf8]{inputenc}

\usepackage{orcidlink}

\begin{document}

%\preprint{-}

\title{
Terahertz radiation induced attractive-repulsive Fermi polaron conversion\\
in transition metal dichalcogenide monolayers
}

\author{A.M. Shentsev\orcidlink{0009-0003-7426-3479}}
\affiliation{Moscow Institute of Physics and Technology, Dolgoprudny, Russia}
\affiliation{L. D. Landau Institute for Theoretical Physics, 142432 Chernogolovka, Russia}
\author{M.M. Glazov\orcidlink{0000-0003-4462-0749}}
\affiliation{Ioffe Institute, 194021 St. Petersburg, Russia }%

\date{\today}

\begin{abstract} 
We present a theoretical study of terahertz radiation-induced transitions between attractive and repulsive Fermi polaron states in monolayers of transition metal dichalcogenides. Going beyond the simple few-particle trion picture, we develop a many-body description that explicitly accounts for correlations with the Fermi sea of resident charge carriers. We calculate the rate of the direct optical conversion process which has a threshold where the terahertz photon energy equals to the Fermi polaron binding energy. This process features a characteristic frequency dependence near the threshold, due to final-state electron-exciton scattering related to the trion correlation with the Fermi sea hole. Furthermore, we demonstrate that intense terahertz pulses can significantly heat the electron gas via Drude absorption enabling an additional, indirect conversion mechanism through collisions between hot electrons and polarons, which exhibits a strong exponential dependence on the electron temperature. Our results reveal the important role of many-body correlations and thermal effects in the terahertz-driven dynamics of excitonic complexes in two-dimensional semiconductors.
\end{abstract}

\maketitle

\section{\label{sec:intro}Introduction}
The optics of semiconductors is largely determined by various Coulomb complexes~\cite{ivchenko2005optical, haug2009quantum, klingshirn2012semiconductor, Semina_2022}. This is especially evident in  atomically thin layers of transition metal dichalcogenides (TMDC)~\cite{Durnev:2018,RevModPhys.90.021001}, where the neutral exciton binding energy is almost two orders of magnitude greater than that of bulk semiconductors  and reaches several hundreds of meV, that makes it stable over a wide temperature range. Relative simplicity of doping atom-thin semiconductors gives rise to a number of manybody states resulting from interaction of excitons with resident charge carriers. It makes  two-dimensional (2D) TMDCs a versatile a versatile platform to study excitonic mixtures with charge carriers~\cite{bruun}.

The simplest possible approach to study manybody effects is to use a few-particle picture. In this approach the response is usually studied in the framework of three-particle complexes, the charged excitons or trions, which are bound states of an exciton with an electron in the conduction band ($X^-$-trion) or with a hole from the valence band ($X^+$-trion), with a binding energy of about 20\ldots 30~meV are observed~\cite{Mak:2013lh,Courtade:2017a}. This, together with the direct-band structure of monolayers of TMDCs, allows one to study pronounced manifestations of excitons and trions in optical spectra. Strong Coulomb interaction provides straightforward optical access to excited states of trions~\cite{PhysRevLett.123.167401,doi:10.1063/5.0013092,PhysRevLett.125.267401,Lin:2022aa} which are hard to observe in  quasi-two-dimensional systems based on conventional semiconductors~\cite{PhysRevB.71.201312,8181-jzj5}. More complex few particle states are also studied experimentally and theoretically~\cite{PhysRevLett.129.076801,ddn8-d8bs}.

Interestingly, the trion binding energies in (2D) TMDCs lie in the terahertz (THz) range of spectra which is actively studied nowadays for fundamental reasons and because of potential applications~\cite{ganichev2006intense,Lampin2020,helm2023elbe,Lu2024}. THz and microwave radiation turns out to be particularly well-suited for studying transitions between electronic and excitonic states in semiconductors~~\cite{ganichev2006intense,leinss2008terahertz, venanzi2021terahertz,Ashkinadze:1988aa,PhysRevB.53.10921,Lifshitz1993}. Recent experimental work~\cite{venanzi2024ultrafast} has demonstrated efficient THz-radiation induced conversion between the neutral and charged  excitons in TMDC monolayers (MLs) and shown the possibility of manipulating the ratio between exciton and trion populations using short, picosecond pulses of terahertz radiation. The observed transitions have been interpreted as THz-induced decomposition of a trion to a neutral exciton and free electron.

\begin{figure}[t]
    \centering
    \includegraphics[width=0.9\linewidth]{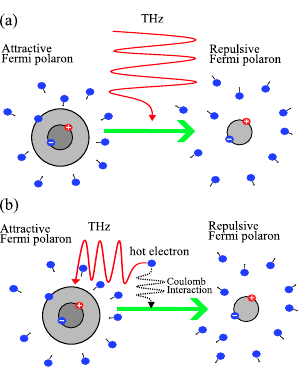}
    \caption{Sketch of studied processes. (a) Transition between attractive and repulsive Fermi polarity due to absorption of THz radiation (Sec.~\ref{sec:ETconv}). (b) Decay of the attractive Fermi polaron state due to interaction with a ``hot'' electron (Sec.~\ref{sec:HEG}).}
    \label{fig0}
\end{figure}

While this few-particle picture provides reasonable description of experiments, it is important to address a role of manybody effects that arise in the case of excitons interacting with resident charge carriers. Indeed, from a manybody perspective, the optical and transport manifestations of trions can be described within the Fermi polaron/Suris tetron approach~\cite{bruun,suris:correlation,PhysRevLett.112.147402,PhysRevB.63.235310,Sidler:2016aa,PhysRevB.95.035417,PhysRevB.102.085304,PhysRevLett.131.106901,PhysRevB.108.125406}, where the correlations between the exciton and Fermi sea are explicitly taken into account. Naturally, at finite density of charge carriers, exciton can bind with any of them that calls for intrinsically manybody approach. While in many cases the trion and polaron pictures provide essentially the same results~\cite{Glazov:2020wf,PhysRevB.105.075311}, the trion and Fermi polaron energy spectrum fine structure related to the electron-hole exchange interaction turns out to be drastically different as suggested theoretically~\cite{iakovlev2023fermi,iakovlev2024longitudinal} and demonstrated experimentally~\cite{yagodkin2024excitonslargepseudomagneticfields}.

%and is formed by exciton correlations with electrons from the sea farm, allowing the binding energy to be tuned by changing the Fermi energy~\cite{iakovlev2023fermi,iakovlev2024longitudinal}. Which, together with valley and spin degrees of freedom~\cite{xu2014spin, lundt2019optical, mak2012control}, provides a lot of scope for experiments.

In this paper, we present a theoretical description of the conversion between the attractive (trion-like) and repulsive (exciton-like) Fermi polarons by the action of terahertz radiation in TMDC MLs, taking into account the correlations of the excitons with the Fermi sea. We show that considering the correlations with the Fermi sea significantly affects the conversion rate at photon energies near the threshold determined by the trion binding energy. In addition to the absorption of light related to the attractive-repulsive polaron (trion-exciton) transition, see Fig.~\ref{fig0}(a), the terahertz pulse heats the electron gas~\cite{ganichev2006intense,venanzi2021terahertz}. Thus, at high intensities or low THz photon energies, it is also necessary to take into account the decay of the trion state due to interaction with high-energy electrons, Fig.~\ref{fig0}(b). We provide estimates for the conversion rates related to the direct interpolaron  and indirect, heating-induced, transitions and highlight specific features of these processes.

The paper is organized as follows: in Sec.~\ref{sec:ETconv} we calculate the transition rate between the attractive and repulsive branches of the Fermi polaron 
due to absorption of THz photons. The effects of the accompanying heating of the electron gas are studied in Sec.~\ref{sec:HEG}, where we first consider the intraband THz absorption at various scattering mechanisms (Sec.~\ref{sec:HET}) and analyze the indirect process of attractive Fermi polaron decay via collisions with high-energy electrons in Sec.~\ref{sec:E-Tr}. The main results and conclusions are summarized in Sec.~\ref{sec:concl}. Additional technical details are provided in the Appendix~\ref{app:details}.

\section{\label{sec:ETconv} THz-absorption induced attractive-repulsive polaron conversion }

Following experimental setting of Ref.~\cite{venanzi2024ultrafast} let us consider excitons in the presence of the Fermi sea in Mo-based TMDC MLs. In such systems, only the intervalley trion, where an exciton in the $K^+$ or $K^-$ valley is bound to a resident charge carrier in the opposite ($K^-$ or $K^+$) valley, is relevant. It is because of the Pauli principle, which makes the intravalley trion with two electrons with the same spin components unstable~\cite{Courtade:2017a}. Thus, following Refs.~\cite{suris:correlation,Schmidt:2018aa,Glazov:2020wf,PhysRevB.95.035417} we consider a simple Hamiltonian
\begin{multline}
\label{Ham}
    \hat{H}_{0} = \sum_{\bm k} \varepsilon_{\bm k}a_{\bm k}^{\dagger}a_{\bm k} + \sum_{\bm k} \varepsilon^X_{\bm k}b_{\bm k}^{\dagger}b_{\bm k}  \\
    + V \sum_{\bm k, \bm k', \bm p, \bm p'}\delta_{\bm k+ \bm p, \bm k' + \bm p'}a^{\dagger}_{\bm p'}b_{\bm k'}^{\dagger}b_{\bm k}a_{\bm p},
\end{multline}
that describes exciton interaction with resident electrons of the opposite valley, e.g., exciton in $K^+$ with electrons in $K^-$ and treats excitons as rigid particles, without considering their internal structure. Here $a^{\dagger}_{\bm k}, a_{\bm k}$ are the creation and annihilation operators of electrons in $K^-$ valley with $\bm k$ being the wavevector reckoned from the $K^-$ point, $b^{\dagger}_{\bm k}, b_{\bm k}$ are the same operators for excitons in the $K^+$ valley, $\varepsilon_{\bm k} = \hbar^2k^2/(2M_e)$ and $\varepsilon^{X}_{\bm k} = \hbar^2k^2/(2M_X)$  are kinetic energies of electrons and excitons, with their effective masses $M_e$ and $M_X$, respectively. The parameter $V < 0$ describes the attractive exciton-electron interaction. We assume that $V$ is independent of the transferred momentum, which corresponds to the $\delta$-function approximation for the exciton-electron interaction potential in the real space which is known to be a reasonable approximation of the main, exchange interaction induced contribution to the exciton-electron binding~\cite{suris:correlation}; see Ref.~\cite{2019arXiv191204873F} for more sophisticated forms of exciton-electron interaction. The interaction of the $K^-$ exciton with $K^+$ electrons is described by the same Hamiltonian. The Hamiltonian~\eqref{Ham} can also be used to describe the interaction of excitons with resident holes in the case of $p$-doped MLs both in W- and Mo-based structures. Note that the Hamiltonian~\eqref{Ham} leads to the high-energy ``ultraviolet'' divergencies. In this work we treat it in a standard way by cutting the momentum integrals at the energies corresponding to the exciton binding energy and expressing the results via the trion binding energy, see Appendix~\ref{app:details} and Refs. ~\cite{suris:correlation,Glazov:2020wf}, we also refer to Ref.~\cite{bruun} for detailed discussion of more advanced approaches.

The interaction of electrons with an electromagnetic field in the electric-dipole approximation is described by the standard perturbation Hamiltonian~\cite{ivchenko2005optical}: 
\begin{equation}
\label{Hle}
    \hat{H}_{l-e} = -\frac{e}{cM_e}(\hat{\bm p}\cdot\bm A) = - \frac{e\hbar}{cM_e}\sum_{\bm k}(\bm k \cdot \bm A)\hat{a}^{\dagger}_{\bm k}\hat a_{\bm k},
\end{equation}
where $\hat{\bm p}$ is the electron quasi-momentum operator, $\hat{\bm p}/M_e$ is the intraband velocity operator of the electron, and the vector potential $\bm A$ of the electromagnetic field is assumed to be coordinate-independent in line with the dipole approximation. The second equality in Eq.~\eqref{Hle} corresponds to the second quantization representation. We assume that the electromagnetic field is classical and monochromatic; $\bm A(t) = \bm A_0 \exp{(-\mathrm i \omega t)} +{\rm c.c.}$, the frequency $\omega$ is in the THz range. 

To study the transitions induced by the interaction~\eqref{Hle} we use the ansatz form of the Fermi polaron wavefunction~\cite{suris:correlation,PhysRevA.74.063628}
\begin{equation}
\label{Psi:k}
    |\Psi_{\bm k} \rangle = \varphi_{\bm k} b_{\bm k}^{\dagger}|FS\rangle + \sum_{\bm p, \bm q}F_{\bm k}(\bm p, \bm q)b_{\bm k-\bm p +\bm q}^{\dagger}a_{\bm p}^{\dagger}a_{\bm q}|FS\rangle,
\end{equation}
where $|FS\rangle$ is the unperturbed Fermi sea of resident electrons, $\bm k$ is the quasi wave vector of the polaron, the coefficients  $\varphi_{\bm k}$ and $F_{\bm k}(\bm p, \bm q)$ describe the contributions of bare exciton and exciton with excited electron-hole pairs to the many particle state, respectively. These coefficients can be found from the Schr\"odinger equation 
\begin{equation}
    \label{schroed:1}
    \hat H_{0}|\Psi_{\bm k}\rangle = E_{\bm k}^{FP} |\Psi_{\bm k}\rangle,
\end{equation} 
where $E_{\bm k}^{FP}$ is the Fermi polaron dispersion.
Hereafter we follow the convention of Refs.~\cite{iakovlev2023fermi,iakovlev2024longitudinal} where the wave vectors $\bm p$ correspond to the states above the Fermi surface ($p>k_F$ with $k_F$ being the Fermi wave vector) and $\bm q$ correspond to the states below the Fermi surface ($q<k_F$). We consider zero-temperature case and assume that $E_T \gg E_F$ and, strictly speaking, $\ln(E_T/E_F) \gg 1$, where $E_T$ is the trion binding energy, which allows us to neglect the dynamics of Fermi sea holes. Equation~\eqref{schroed:1} provides two solutions: repulsive and attractive Fermi polarons, that in the limit of $k_F \to 0$ reduce to the free exciton and trion, respectively.

Qualitatively, in the trion approach the photon absorption removes the electron from the trion making it to dissociate into the free exciton and electron. In the Fermi polaron approach the light-matter interaction transfers the attractive Fermi polaron state~\eqref{Psi:k} to the repulsive polaron state with an extra electron-hole pair in the Fermi sea, that is exciton plus electron plus Fermi sea hole.\footnote{In the absence of additional scattering processes by impurities or phonons the Hamiltonian~\eqref{Hle} does not provide a transfer between the attractive and repulsive states in the form of Eq.~\eqref{Psi:k} because THz-created electron-hole pair in the Fermi sea has zero net momentum and the density of final states is, hence, vanishingly small.} Such a continuum state is described by the wavefunction in the form similar to Eq.~\eqref{Psi:k}
\begin{multline}
\label{Phi:k}
    |\Phi_{\bm k,\bm p, \bm q} \rangle =  b_{\bm k-\bm p + \bm q}^{\dagger}a_{\bm p}^{\dagger}a_{\bm q}|FS\rangle\\
    + \sum_{\bm p'}U_{\bm k, \bm p, \bm q}(\bm p')b_{\bm k-\bm p' +\bm q}^{\dagger}a_{\bm p'}^{\dagger}a_{\bm q}|FS\rangle,
\end{multline}
where the function $U_{\bm k, \bm p, \bm q}(\bm p')$ takes into account exciton-electron scattering which can be found from the Schr\"odinger equation in the form similar to Eq.~\eqref{schroed:1} As before, we assume that $\bm q$ is below the Fermi surface and disregard the exciton-hole scattering.
%are found from self-consistency equations $H|\psi\rangle = E_{\psi}|\psi\rangle~\cite{iakovlev2023fermi}$, it is also assumed in this and following equations that $ p,p_0 > k_F$, while $ q,q_0 < k_F$. In this approximation we assume temperature equal to zero and carry out assessments according to the parameter of smallness  $1/\ln(E_T/E_F) \ll 1$, therefore we neglect the scattering of holes $\bm q_0$ in the Fermi sea and terms with a large number of fermion creation-annihilation operators. 
%
The explicit expressions for the functions $\varphi_{\bm k}$, $F_{\bm k}(\bm p, \bm q)$, and $U_{\bm k, \bm p, \bm q}(\bm p')$ as well as the technical details are presented in Appendix~\ref{app:details}.

\begin{figure}[t]
    \centering
    \includegraphics[width=\linewidth]{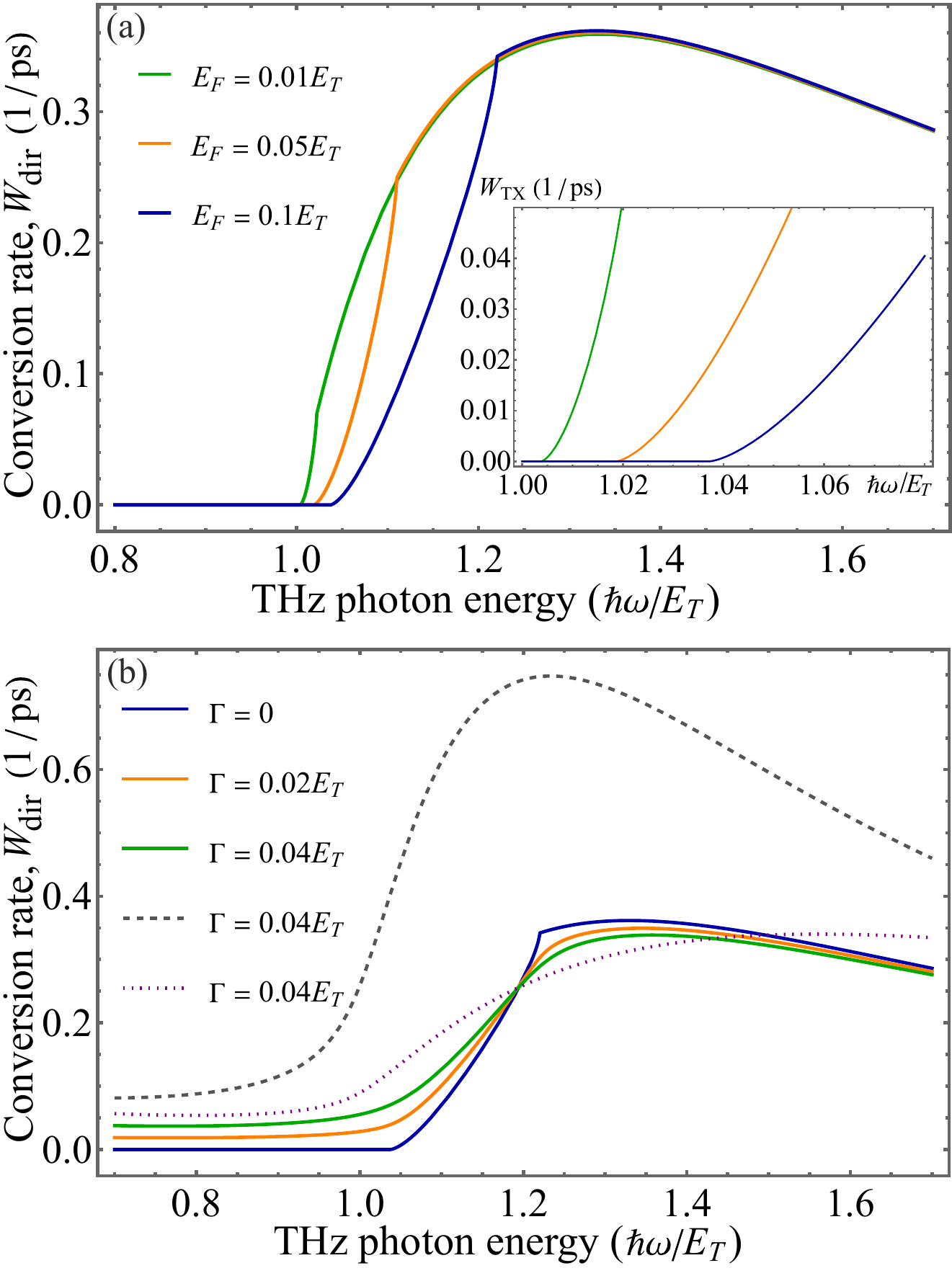}
    \caption{(a) Dependence of the transition rate $W_{\rm dir}(\omega)$ on the frequency of terahertz radiation at different Fermi energies calculated disregarding the broadening. Inset shows asymptotic of $W_{\rm dir}(\omega)$ near the threshold $\hbar\omega - |E_{FP}| \ll E_F \propto (\hbar\omega - |E_{FP}|)^{3/2}$. (b) Solid lines show the transition rate $W_{\rm dir}(\omega)$ at $E_F = 0.1 E_T$, for different spectral broadening $\Gamma$. The parameters of calculation~\cite{venanzi2024ultrafast, naftaly2011hexagonal}: $E_T = 25$ meV, $I = 1.25$~$\mu$J/(cm$^2$ps), $M_e = 0.5~M_0$, $M_x = 1.1~ M_0$, where $M_0$ is free electron mass. The dashed curve is calculated in the trion approach of~\cite{venanzi2024ultrafast}, with the exciton radius $a = 1$ nm and the trion radius $b = 3$ nm, the dotted curve is the same model with $a = 0.6$ nm, $b = 1.1$ nm.}
    \label{fig1}
\end{figure}

 Matrix element of the THz-radiation induced transition from the state $|\Psi_{\bm k}\rangle$ to the state $|\Phi_{\bm k, \bm p, \bm q}\rangle$ takes the form
\begin{multline}
\label{me}
    \Upsilon_{\bm k}( \bm p, \bm q) = \langle\Phi_{\bm k, \bm p, \bm q}|\hat{H}_{l-e} |\Psi_{\bm k}\rangle =  \\
    = -\frac{e\hbar}{cM_e}F_{\bm k}(\bm p, \bm q) (\bm A \cdot\bm p)\\   - \frac{e\hbar}{cM_e}\sum_{\bm p'} F(\bm p', \bm q)U^{*}_{\bm k, \bm p, \bm q}(\bm p') (\bm A \cdot\bm p').
\end{multline} 
Note that the second-to-last line in Eq.~\eqref{me} describes the transition from the attractive polaron to the free exciton state and the second term takes into account the modification of the matrix element due to the electron-exciton scattering (i.e., repulsive Fermi polaron effect). In what follows we consider the conversion processes for polarons with small wavevectors. Thus we set $\bm k=0$ and derive the closed-form expression for the transition element $\Upsilon_0( \bm p, \bm q)$  in the leading order in $1/\ln(E_T/E_F)$ as
\begin{multline}
    \label{M:TX}
    \Upsilon_{0}( \bm p, \bm q) = -\frac{e}{cM_e}\frac{\left[\bm A\cdot\left(\bm p - \frac{M_e}{M_T}\bm q\right)\right]}{2\sqrt{\frac{E_T}{E_F}\left(\frac{M_T}{M_X}\right)^3}\sinh\left[\frac{1}{2}\left(\frac{M_X}{M_T}\right)^2\right]}\\
    \times\frac{2\pi\hbar^{3}}{\mu S}\frac{E_T(E_T + E_{FP}+\varepsilon_{\bm q}-\varepsilon_{\bm q}^T-\frac{M_T}{M_x}E_F)^{-1}}{(E_{FP}-\varepsilon^X_{-\bm p + \bm q}-\varepsilon_{\bm p}+\varepsilon_{\bm q})},
\end{multline}
where $M_T = M_e + M_X$ is the trion mass, $\mu = M_eM_X/M_T$ is the reduced mass of the electron-exciton pair, $\varepsilon_k^T = \hbar^2k^2/2M_T$ is the kinetic energy of trions, $S$ is the normalization area of the system, and $E_{FP} \equiv E_{0}^{FP}< 0$ is the attractive Fermi polaron energy at $\bm k=0$, which for the case of Mo-based monolayers is equal to~\cite{iakovlev2023fermi}
\begin{equation}
   E_{FP} = - E_T + E_F\left(\frac{M_T}{M_X}- \frac{M_X/M_T}{1-e^{-(M_X/M_T)^2}}\right).
\end{equation} 
Note that the absolute value of the attractive Fermi polaron (Suris tetron) energy differs from the trion binding energy $E_T$ by the amount proportional to the electron Fermi energy.
Then, the rate of the THz-induced transitions is found using Eq.~\eqref{M:TX} and Fermi's golden rule as
\begin{multline}
    \label{W:TX}
    W_{\rm dir} = \frac{2\pi}{\hbar}\sum_{\bm p, \bm q}|\Upsilon_0(\bm p,\bm q)|^2\\\delta(E_{FP}+\hbar\omega - \varepsilon^X_{-\bm p + \bm q}-\varepsilon_{\bm p}+\varepsilon_{\bm q}).
\end{multline} 

The results of numerical calculation of the transition rate for different values of the electron Fermi energy with respect to the trion binding energy are shown in Fig.~\ref{fig1}(a). The dependence of $ W_{\rm dir}$ on the THz frequency shows a threshold where $\hbar\omega \approx -E_{FP}\approx E_T$ since for lower energies the transitions are forbidden by the conservation laws, reaches a maximum and drops as $\hbar\omega$ increases. 

To gain further insight and analyze the THz photon energy dependence of $W_{\rm dir}$ in more detail we derive from Eq.~\eqref{W:TX} explicit analytical expressions for $W_{\rm dir}$ in the following two cases. The first case corresponds to the onset of the THz absorption spectrum, i.e., the frequency range $0<\hbar\omega - |E_{FP}|\ll E_F$, where the Fermi-polaron effects are particularly important:
\begin{multline}
    \label{W:beg}
    W_{\rm dir} \approx I\times \frac{8\pi\alpha e^{(M_X/M_T)^2}\hbar^2}{3}\frac{\sqrt{M_X^3/M_e^3}}{M_T}\\
    \times\frac{E_T(\hbar\omega-|E_{FP}|)^{3/2}}{(\hbar\omega)^4\sqrt{E_F}} \theta(\hbar\omega - |E_{FP}|).
\end{multline}
Here  $\alpha = 1/137$ is the fine structure constant, $I$ is the radiation intensity on the sample (for simplicity we disregard the dielectric contrast between the sample and surroundings) The main result of Eq.~\eqref{W:beg} is the small-frequency asymptotics $\propto(\hbar\omega - |E_{FP}|)^{3/2}$, see, in particular, the inset to Fig.~\ref{fig1}(a). Note that taking into account the terms which are small in the parameter $1/\ln(E_T/E_F)$, leads to a change in the prefactor in Eq.~\eqref{W:beg} but the power law remains the same. In the vicinity of the threshold, the increase in $W_{\rm dir}$ with increase in $\hbar\omega$ is related to the fact that for larger THz frequencies a larger range of available hole states in the Fermi sea. At $\hbar\omega-|E_{FP}| = 4(M_e/M_x)E_F$ all electrons from the Fermi sea are involved in the THz-induced transition and a kink in the curves occurs in Fig.~\ref{fig1}(a) since there are no more electrons to add. With further increase in $\hbar\omega$ where $\hbar\omega > -E_{FP} + 4(M_e/M_x)E_F$, we turn to the second case where analytical result is derived as
\begin{equation}
    \label{W:TP}
     W_{\rm dir} \approx I\times  \frac{4\alpha \pi^2\mu\hbar^2 E_{T}}{M_e^2}
    \frac{\hbar\omega -|E_{FP}+ \beta E_F|}{(\hbar\omega)^4},
\end{equation}
with
\[
\beta = \frac{M_X}{4M_T}\frac{e^{\left(\frac{M_X}{M_T}\right)^2}-1-\left(\frac{M_X}{M_T}\right)^2}{\sinh^2\left[\frac{1}{2}\left(\frac{M_X}{M_T}\right)^2\right]},
\]
where all curves in Fig.~\ref{fig1}(a) merge. 
%\addSasha{Switching between regimes \eqref{W:beg} and \eqref{W:TP} corresponds to an increase in the initial possible positions of electrons $\bm q$ in the decay states~\eqref{Phi:k}, thus at low energies~\eqref{W:beg} there is a decay with an electron-hole pair near the Fermi surface with $|\bm p| \approx k_F \approx |\bm q|$; at $\hbar\omega-|E_{FP}| = 4(M_e/M_x)E_F$ all electrons from the Fermi sea are involved and a kink in the curves occurs in Fig.\ref{fig1} (a).} 
This regime corresponds basically to the trion result of Ref.~\cite{venanzi2024ultrafast}. The transition rate has a maximum that is weakly dependent on the Fermi energy at $\hbar\omega = 4|E_{FP}+ \beta E_F|/3 \approx 4E_T/3$. The peak position and decay asymptotics at $\hbar\omega \gg E_T$ depend on the shape of the electron-exciton relative motion envelope function, see below for more detailed comparison.  For the evaluation of Eq.~\eqref{W:TX} in the general case see Appendix~\ref{app:details}.

Short-range static disorder and interaction with phonons at finite temperature gives rise to the scattering processes which eventually result in the decay of the quasiparticles, while the long-range disorder results in the inhomogeneous broadening of the resonances. To illustrate the effect we introduce a non-zero broadening in the energy conservation $\delta$-function in the Fermi's golden rule~\eqref{W:TX} replacing it with the Lorentzian with the result
\begin{multline}
    \label{W:FT}
    W_{\rm dir} = \frac{2\pi}{\hbar}\sum_{\bm p, \bm q}|\Upsilon_0(p,q)|^2\\
    \frac{1}{\pi}\frac{\Gamma}{(E_{FP}+\hbar\omega - \varepsilon^X_{-\bm p + \bm q}-\varepsilon_{\bm p}+\varepsilon_{\bm q})^2 + \Gamma^2}.
\end{multline}
Here $\Gamma$ is the effective linewidth (typically in meV range~\cite{wagner:trions,PhysRevB.108.125406,venanzi2024ultrafast}). For $\Gamma \ll E_F$ the modification of the polaron wavefunctions can be neglected in the transition matrix element $\Upsilon_0(p,q)$. Naturally, in the case of $\Gamma \sim E_F$ the difference between calculations with or without taking into account correlations with the Fermi sea is insignificant~\cite{Glazov:2020wf}. The results of calculations by Eq.~\eqref{W:FT} are shown in Fig.~\ref{fig1}(b). The main effect of the scattering is in the vicinity of the threshold energy: naturally, broadening smooths-out the threshold and slightly reduces the transition rate for $\hbar\omega> E_T$. 

For comparison dashed and dotted curves in Fig.~\ref{fig2}(b) show the results of the trion approach developed in Ref.~\cite{venanzi2024ultrafast}. In that case electron-exciton correlations in the final state and the correlations of the trion with a hole in the Fermi sea are disregarded. Moreover, Ref.~\cite{venanzi2024ultrafast} uses well established exponential form of the relative motion wavefunction (see Ref.~\cite{Semina_2022} for details on variational approach to the trion problem). For the experimentally relevant parameters (dashed curve) the transition rate is about twice larger than our result. The analysis shows that it is mainly the effect of the relative motion wavefunction shape: modified Bessel function in our Fermi-polaron approach vs. exponent in the trion approach. Adjusting the exciton and trion radii (dotted curve) we can obtain reasonable agreement in the magnitude of the transition rate, but the shape differs both in the vicinity of THz-absorption onset and at large frequencies. It is related again to the shape of the relative motion wavefunction  which leads to different Fourier transforms and, consequently, different wavevector dependence of the matrix element $\Upsilon_0(p,q)$ and, eventually, the transition rate spectral dependence, see Eq.~\eqref{Wdir:exp} and \ref{app:details} for details.

\section{\label{sec:HEG} Effect of electron gas heating by THz radiation}

%\section{\label{sec:HEG} Heating of electron gas under the influence of THz radiation}

The conversion rate calculated above is proportional to the intensity of electromagnetic radiation and has a maximum at a photon energy of $\hbar\omega \approx 4E_{tr}/3$. To achieve high transition rates (see Fig.~\ref{fig1}), several picosecond long, high-fluence $\phi \sim 1$~$\mu$J/cm$^2$ terahertz pulses can be used~\cite{venanzi2024ultrafast}, which corresponds to a radiation intensity of the order of $\mu$J/(cm$^2$ps). As a result, a major part of the attractive polarons (trions) can be converted to repulsive polarons (excitons). On the other hand, with such pumping parameters, the heating of the electron gas caused by the THz absorption can be significant. The electron temperatures can reach several tens of Kelvin, which corresponds to the average kinetic energy $k_BT \sim 0.1-0.3~E_T$~\cite{venanzi2021terahertz}. In this section we provide an analytical model for the electron gas heating under the absorption of THz radiation and, consequently, calculate the additional contribution to the attractive-repulsive polaron conversion related to the collisions of polarons with high-energy electrons.

\subsection{\label{sec:HET}Electron gas heating}

The electron gas heating by THz radiation  is mainly related to the Drude absorption processes where electrons absorb photons and scatter by defects or phonons. To describe the process, we use the second order perturbation theory in the high-frequency field approximation $\omega \tau \gg 1$, where $\tau$ is the electron scattering time on the order of tenths to units of picoseconds.\footnote{For typical parameters estimates show that $\omega\tau \sim 10$.}  In this case the matrix element of the transition from the state $\bm k$ in $\bm k'$ with photon absorption is expressed as follows:
\begin{equation}
\label{Matr:el}
    \Upsilon_{\bm k, \bm k'} = {-}\frac{1}{\hbar\omega}\langle \bm k'| [\hat{H}_{l-e}, \hat{H}_{m}]|\bm k\rangle,
\end{equation}
where $\hbar \omega$ is the photon energy, the light-matter interaction Hamiltonian is presented in Eq.~\eqref{Hle} and $\hat{H}_{m}$ is the Hamiltonian that describes the scattering, i.e., the processes that do not conserve the momentum of the electron gas.
%, as in scattering by phonons or defects. 

\subsubsection{\label{sec:HEG:PD} Scattering by point defects} 

First, let us consider scattering on the static short-range defects. Then, in the second quantized form, the interaction Hamiltonian reads:
\begin{equation}
\label{Def}
    \hat{H}_m \equiv\hat{H}_{D} = \frac{u}{S}\sum_{i,\bm k, \bm k'}e^{i(\bm k -\bm k')\bm R_i} \hat{a}^{\dagger}_{\bm k'}\hat{a}_{\bm k}.
\end{equation}
Here $\bm R_i$ is the position of the $i$th defect, and $u$ is the effective strength of the defects' potential. We assume that all defects are identical. In this approximation, the matrix element~\eqref{Matr:el} has the following form:
\begin{equation}
    \Upsilon^{D}_{\bm k, \bm k'} = \frac{eu (\bm k' -\bm k)\cdot \bm A}{ c\omega SM_e} \sum_{i}e^{i(\bm k -\bm k')\bm R_i}.
\end{equation}
Neglecting an interference at scattering off different defects we get
\begin{equation}
\label{U:Def}
    |\Upsilon^{D}_{\bm k, \bm k'}|^2 \approx \frac{n}{S}\left(\frac{eu (\bm k' -\bm k)\cdot \bm A}{ c M_e\omega}\right)^2,
\end{equation}
where $n$ is the density of defects. The absorption power density (i.e., the rate of the electron gas heating since the electron scattering is elastic) is given by~\cite{Budkin:2011aa}):
\begin{multline}
    Q_D = \hbar\omega \times 2\times\frac{2\pi}{\hbar}\frac{1}{S}\sum_{\bm k, \bm k'}|\Upsilon^{D}_{\bm k, \bm k'}|^2\\
    (f_{\bm k}-f_{\bm k'})\delta(\varepsilon_{\bm k'} -\varepsilon_{\bm k }-\hbar\omega).
\end{multline}
Here $f_{\bm k}$ is equilibrium Fermi-Dirac distribution function, and the extra factor $2$ accounts for the spin/valley degeneracy of electrons. In the experimentally relevant situation $\hbar\omega \sim E_T\gg E_F$ we have 
\begin{equation}
\label{Q:D}
    Q_D = \frac{4\pi}{c}I\times\frac{n_ee^2}{2\tau\omega^2M_e}
    \times \left(1 + \frac{2k_B T}{\hbar \omega}\right) (1-e^{-\frac{\hbar\omega}{k_BT}}),
    %{\times \frac{(\hbar\omega+2k_BT)(1-e^{-\frac{\hbar\omega}{k_BT}})}{\hbar\omega}},
\end{equation}
where $n_e$ is the electron gas density, $\tau^{-1} = nu^2M_e/\hbar^3$ is the scattering time, $T$ is the electron gas temperature. Taking temperature into account is only significant when $k_BT\sim\hbar\omega\gg E_F$, hence, we assumed the Boltzmann statistics of the distribution function to calculate the temperature-related contribution. Note that in the classical case where $\omega\tau \gg 1$ and $k_BT \ll E_F$ but $\hbar\omega\ll E_F$ the heating rate is twice larger than that given by Eq.~\eqref{Q:D}.
%Main contrasts from the classic $Q = \frac{4\pi}{c} t^2I \cdot\frac{n_ee^2}{\tau\omega^2M_e}$ are in the coefficient $1/2$ arising from the limit $\hbar\omega \gg E_F$.  

\subsubsection{\label{sec:HEG:LAP} Scattering by long-wavelength acoustic and optical phonons} 

Second, let us consider the scattering of electrons by long-wavelength acoustic and optical phonons. In transition metal dichalcogenide monolayers, they mainly determine the scattering time~\cite{kaasbjerg2012phonon,kaasbjerg2013acoustic} and thermal~\cite{kaasbjerg2014hot, venanzi2021terahertz} exchange of electrons with the lattice. The electron-phonon interactio Hamiltonian reads
%The most important scattering mechanism is that of the deformation potential (the piezoelectric coupling is weak in two-dimensional systems):
%The characteristic energy of a photon corresponding to a change in the momentum of an electron is equal to $\hbar s\sqrt{2M_eE_{T}/\hbar^2} \sim meV$ and is in the long-wavelength limit. Thus the deformation potential:
\begin{multline}
\label{eq:defpot}
    \hat{H}_{DP} = i\times\sum_{\bm q,\bm k} \left(\frac{\hbar}{2\rho \omega_{\lambda}(\bm{q})S}\right)^{1/2}M_{\lambda}(\bm q)\times\\(\hat{b}_{\lambda,\bm q}\hat{a}^{\dagger}_{\bm k+\bm q}\hat{a}_{\bm k} +\hat{b}_{\lambda,\bm q}\hat{a}^{\dagger}_{\bm k-\bm q}\hat{a}_{\bm k}),
\end{multline} 
where $\lambda$ denotes the phonon branch, $\omega_\lambda(\bm q)$ is its dispersion, $M_\lambda(\bm q)$ is the matrix element, $\hat{b}_{\lambda,\bm q}, \hat{b}_{\lambda,\bm q}^\dag$ are the phonon field operators, and $\rho$ is the two-dimensional mass density of the crystal. Hereafter we use simplified approach suggested in Ref.~\cite{PhysRevB.90.045422} where for interaction with acoustic phonons we take into account the deformation potential (since the piezoelectric coupling is weak in two-dimensional systems) and consider longitudinal acoustic branch ($\lambda=LA$) with $\omega_{LA} = sq$ ($s$ is the speed of sound) and $M_{LA}(\bm q) = \Xi q$ ($\Xi$ is the deformation potential) while for interaction with optical phonons we use the effective optical phonon model with a single branch ($\lambda=O$) and fixed frequency $\omega_O$ interaction matrix element $M_O(\bm q) = D_0$, see~\cite{PhysRevB.90.045422} for details.
The phonon-assisted transition matrix element~\eqref{Matr:el} has the form 
\begin{equation}
    \Upsilon_{\bm k, \bm k'}^{\lambda} = -i\frac{e}{cM_e}\left(\frac{\hbar |M_{\lambda}(\bm q)|^2}{2\rho \omega_{\lambda}(\bm q) S}\right)^{1/2}\frac{(\bm k - \bm k')\cdot\bm A}{\omega}.
\end{equation}

As we will see below, processes that take into account long-wavelength acoustic phonons are important in the regions of relatively large $\hbar\omega$ and small $T$, therefore, to estimate the absorbed power per unit area, we can write it as follows\footnote{One can check that for THz frequencies relevant acoustic phonon energy $\sim \sqrt{\hbar\omega M_e s^2}\ll \hbar\omega$. That is why electron gains energy mainly from the photon $\varepsilon_{\bm k'} - \varepsilon_{\bm k} \approx \hbar\omega$. Moverover, one can disregard $\varepsilon_{\bm k}$ in the energy conservation law.}
%Since $\hbar \omega$ much larger than the other parameters of the system, we can ignore the difference between processes with emission and absorption of phonons, and similarly to (17) we can neglect the Fermi statistics and the initial energy of the electrons. After these simplifications, the absorption power is written as follows
\begin{multline}
    \label{Q:DP}
    Q_{LA} = \hbar\omega\times\frac{2\pi}{\hbar}n_e\sum_{\bm k'}|\Upsilon_{\bm 0, \bm k'}^{LA}|^2(1 + 2N_{\bm k'})\delta(\varepsilon_{\bm k'}-\hbar\omega)\\
    =\frac{4\pi}{c}I\times\frac{n_ee^2\Xi^2M_e^{1/2}}{\hbar\rho s(2\hbar\omega)^{3/2}}\coth{\left(\frac{s\sqrt{M_e\omega/2\hbar}}{k_BT_{l}}\right)}.
\end{multline}
Here $N_{\bm k}$ is Bose distribution function for LA phonons at the lattice temperature $T_l$. Note that for typical wavevectors of invloved phonons the coefficient $1 + 2N_{\bm k}$ does not differ much from unity.  In contrast to Eq.~\eqref{Q:D}, scattering by phonons gives a $\omega^{-3/2}$ law instead of $\omega^{-2}$ for point defects. It is because the involved phonon energy is, in accordance with the conservation laws $\sqrt{\hbar\omega M_{e} s^2} \gg k_B T_l$. Note that at lower frequencies where $\sqrt{\hbar\omega M_{e} s^2} \ll k_B T_l$ we obtain standard Drude expression for the absorption.\\

%\addSasha{As an optical phonon we consider only the $A_1$ branch for MoSe$_2$. In this case $\hbar\omega \approx \hbar\omega_{A_1} \approx 30.3$~meV, $M_{A_1}(\bm q) = D^0_{A_1}$ is the deformation potential constants for optical $A_1$ phonons for intravalley electron transition~\cite{}. The absorption power has the following form} 
The absorbed power density at the optical phonon scattering can be expressed in the form [cf.~Ref.~\cite{Budkin:2011aa}]
\begin{multline}
    Q_{O} = (\hbar\omega-\hbar\omega_{O}) \times 2\times\frac{2\pi}{\hbar}\frac{1}{S}\sum_{\bm k, \bm k'}|\Upsilon^{O}_{\bm k, \bm k'}|^2\\
    \times f_{\bm k}(1-f_{\bm k'})\delta(\varepsilon_{\bm k'} -\varepsilon_{\bm k }-\hbar\omega+\hbar\omega_{O}).
\end{multline}
Here we assume the lattice temperature $T_l\ll \hbar\omega_{O}$ that allows us to consider only processes accompanied by spontaneous emission of optical phonons. To simplify the evaluation we assume that the electrons are non-degenerate and obtain
\begin{multline}
    \label{Q:DP:A1}
    Q_{O} = \frac{4\pi}{c}I\times\frac{n_ee^2}{2\omega^2M_e}\times\frac{(D_{0})^2M_e}{2\rho\hbar^2\omega_{O}}\\
    \times\frac{(\hbar\omega-\hbar\omega_{O})(|\hbar\omega- \hbar\omega_{O}|+2k_BT)}{(\hbar\omega)^2}\\
    \times\left[1 -\theta(\hbar\omega_{O}-\hbar\omega)\left(1-e^{-\frac{\hbar\omega_{O}-\hbar\omega}{k_BT}}\right)\right].
\end{multline}
A characteristic feature of this process is that the electron-optical phonon interaction may result both in the electron gas heating (at $\hbar\omega >\hbar\omega_O$) and cooling (at $\hbar\omega < \hbar\omega_O$ the absorbed power $Q_O<0$), cf.~Ref.~\cite{Budkin:2011aa}. In the latter case, however, the cooling efficiency is proportional to a fraction of electrons with energies $\varepsilon_{\bm k} \gtrsim \hbar\omega_O - \hbar\omega$. As a result, the  the dependence of the absorbed power on the frequency of THz radiation becomes non-monotonous.

%\sSasha{such that for $\omega \gg\omega_{d-ph}$ the phonon-assisted absorption dominates.
%For MoS$_2$~\cite{kaasbjerg2012phonon,PhysRevB.90.045422,shree2018exciton}: $s = 6.7\cdot 10^3$ m/s, $\rho = 3.1\cdot10^{-6}$ kg/m$^2$, $\Xi = 2.8$ eV and if we take $\tau = 1...10$ ps, then $\omega_{d-ph} \approx 10^{15}...10^{13}~1/\text{s}$.  For the case of interest to us $\omega \sim 1\ldots10$ THz both absorption mechanisms -- via phonon and static defects scattering -- coexist.  
%}

\subsubsection{Electron gas temperature} 

In this work for simplicity we consider electrons described by quasiequilibrium distribution which is parametrized by the electron density $n_e$, which remains constant under THz illumination, and temperature $T$. Actual form of the electron distribution function determined from the full kinetic equation can be different, but for our purposes it is sufficient to determine the electron gas temperature that controls the relevant electron-polaron scattering processes.
To determine the electron gas temperature $T$ under THz irradiation one needs, in general, to solve the heat balance equation
\begin{equation}
    \label{heat:balance}
    \frac{d}{dt}\int_{T_l}^{T(t)} C(T') dT' \equiv C(T) \frac{dT}{dt} = (Q - Q_l),
\end{equation}
where $C(T)$ is electron gas heat heat capacity, $Q=Q_D+Q_{LA}+ Q_{O}$ is the heating rate caused by the Drude absorption, and $Q_l$ is the energy loss rate to the lattice. The heat capacity $C(T)$ varies between two limits
\begin{equation}
    \label{spec:heat}
    C(T) = \begin{cases}
         \frac{\pi^2}{3}n_ek_B\frac{k_BT}{E_F}, \quad k_B T \ll E_F,\\
         n_ek_B, \quad k_B T \gg E_F, 
    \end{cases}
\end{equation}
valid for degenerate and non-degenerate electrons, respectively. To simplify calculations we use a piecewise approximation for $C(T)$ based on two asymptotics in Eq.~\eqref{spec:heat}.

The electron energy is lost as a result of emission of the  acoustic and optical phonons. The energy loss rate can be written in the form
\begin{equation}
    \label{Q:l}
    Q_l =  -C(T)\frac{(T-T_l)}{\tau_{LA}} - R_O,
\end{equation}
where the energy relaxation by acoustic phonons is described by a temperature-independent relaxation time $\tau_{LA}$~\cite{gantmakher87,PhysRevLett.124.166802}
\begin{subequations}
    \label{energy:relaxation}
    \begin{equation}
        \label{tau:e:LA}
        \frac{1}{\tau_{LA}} = \frac{2 M_e^2 |\Xi|^2}{\rho \hbar^3},
    \end{equation}
with $\tau_{LA}\sim 10\ldots 50$~ps~\cite{venanzi2021terahertz},
while the optical phonon emission results in the energy loss in the form [cf. Ref.~\cite{gantmakher87}]
\begin{equation}
    \label{Ro}
    R_O= \frac{|D_0|^2M_e}{2\rho\hbar}n_e e^{-\frac{\hbar\omega_{O}}{k_BT}}.
\end{equation}
\end{subequations}
Note that Eq.~\eqref{Ro} is valid for $T\gg T_l$, but the contribution due to the optical phonon scattering at $T\sim T_l$ for low and moderate lattice temperature to the electron cooling is negligible.

\begin{figure}[h]
    \centering
    \includegraphics[width=\linewidth]{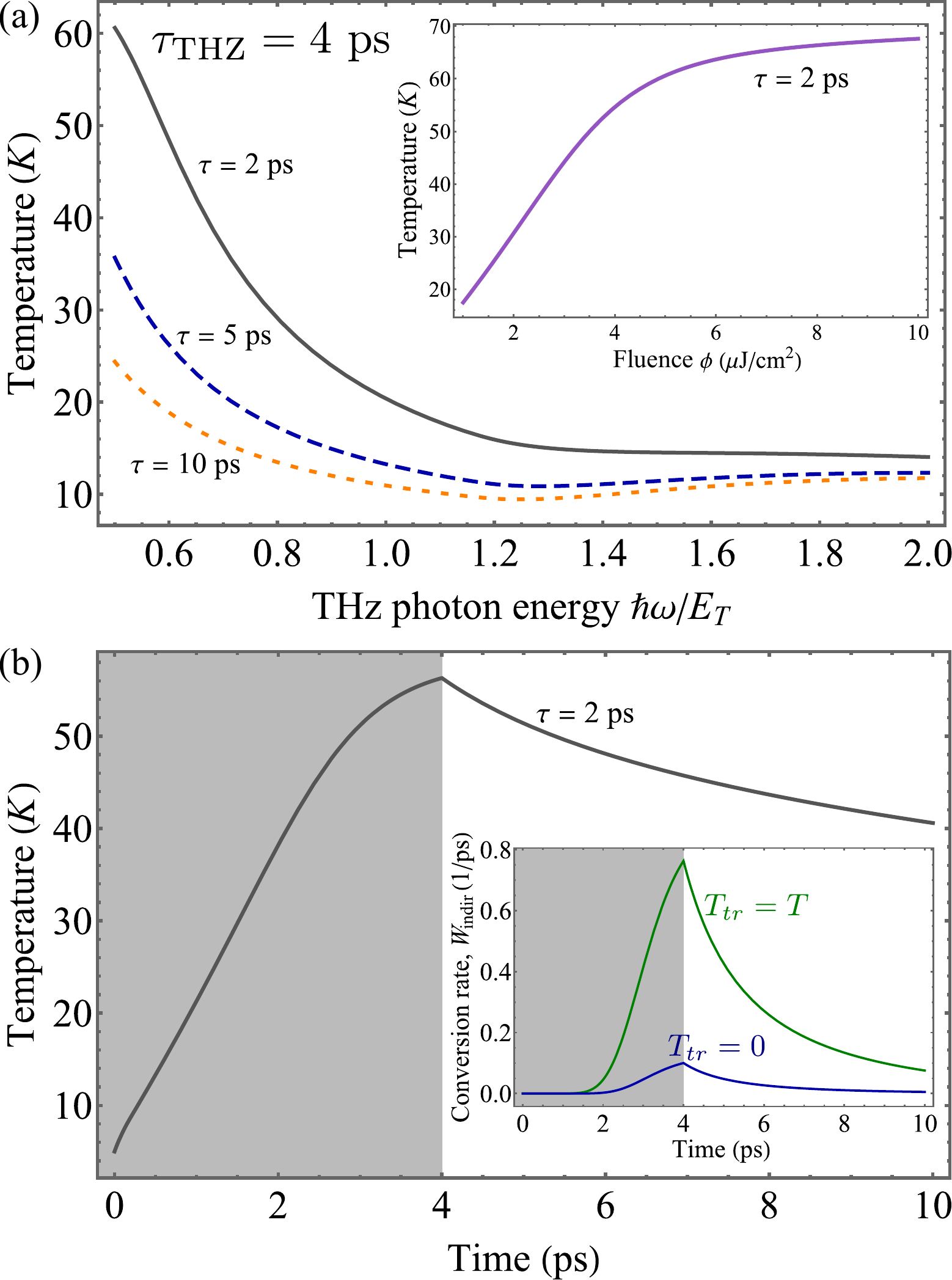}
    \caption{(a) Dependence the electron gas temperature  $T$ right after THz irradiation pulse calculated from Eq.~\eqref{heat:balance} for different scattering time by static impurities $\tau$ as a function of the THz frequency. Inset shows dependence the electron gas temperature  $T$ on the THz pulse fluence $\phi$ with $\tau = 2$~ps and $\hbar\omega = 0.5 E_T$. (b) Time dependence of the electron gas temperature calculated from Eq.~\eqref{heat:balance} for $\tau=2$~ps. Inset shows the conversion rate for the indirect process as a function of time, see text for details. Shaded area shows the duration of the THz pulse. The parameters of calculation: lattice temperature $T_l = 5$ K, pulse duration $\tau_{\rm THz} = 4$ ps, $E_F = 0.1E_T$, $\phi = I\tau_{\rm THz} = 5$~$\mu$J/cm$^2$, $D_0 = 5.2\cdot 10^8$ eV/cm, $\Xi = 3.4$ eV, $\rho = 4.46\cdot 10^{-7}$ g/cm$^2$, $s = 4.1\cdot 10^5 $ cm/s, $\tau_{LA} = 30$~ps, other parameters are the same as ones used for the calculations for Fig.~\ref{fig1}. Here we approximate the electron gas heat capacity as $C(T) \approx C_F^{2D}(T)$ for $k_BT < 3E_F/\pi^2$ and $C(T) \approx C^{2D}_B$ for $T>3E_F/\pi^2$.}
    \label{fig2}
\end{figure}

The electron gas temperature right after the THz pulse calculated by solving Eq.~\eqref{heat:balance} is shown in Figure~\ref{fig2}(a). The analysis shows that the heating is mainly related to the THz photon absorption with scattering by point defects~\eqref{Q:D} and cooling due to the emission of optical phonons, see Eqs.~\eqref{Q:DP:A1} and \eqref{Q:l}, particularly, at low THz phonon energy where the Drude absorption is particularly strong and the temperatures are high. Acoustic phonon scattering becomes significant only at high frequencies of THz radiation, where the electron gas temperature reaches relatively low values of $10\ldots 20$~K. Exponential activation of cooling due to optical phonons leads to saturation-like behavior of the electron temperature with increase in the fluence, see inset in Fig.~\ref{fig2}(a). Figure~\ref{fig2}(b) shows the time dependence of the electron temperature that first increases almost linearly during the pulse action because of the Drude heating, then shows an onset of saturation where cooling by optical and acoustic phonons start to play a role. After the pulse ends (end of the shaded area in Fig~\ref{fig2}(b) the temperature goes down as a result of the cooling processes related to the phonon emission. 

%\sSasha{Under the experimentally conditions of pulsed excitation and under assumptions of the degenerate %electron gas with the heat capacity $$C^{2D}_F(T) = \frac{\pi^2}{3}n_ek_B\frac{k_BT}{E_F}$$ and weak %losses to the lattice we have
%}
%\begin{equation}
%\label{T:ai}
%    \sSasha{T = \sqrt{T^2_{l}+\frac{6Q \tau_{\rm THz} E_F}{\pi^2k_B^2n_e}},}
%\end{equation}
%\sSasha{where $\tau_{\rm THz}$ is the duration of the THz pulse. For the parameters realized in the experiment\cite{venanzi2021terahertz,venanzi2024ultrafast}, the temperature of the electron gas after THz pulse is of the order of several tens of Kelvins, see Fig.~\ref{fig2}. This result is in agreement with Ref.~\cite{venanzi2021terahertz}. Note that Eq.~\eqref{T:ai} becomes inapplicable at $T \gtrsim E_F/3$, in which case the statistics starts to approach the Boltzmann one and the heat capacity can be estimated as $C_B^{2D} = n_ek_B$.
%}
 
\subsection{\label{sec:E-Tr}Conversion due to the electron-polaron collisions}

The electron gas heating results in increased number of the charge carries with elevated energies $\varepsilon_{\bm k}> E_T$. The scattering of attractive polarons by such high-energy electrons can also result in the attractive-repulsive polaron conversion. Such a process can be viewed as a sort of the shake-up process where the excited Fermi sea impacts the polarons. It is also similar to the charge carrier scattering assisted transitions between bound and localized excitons~\cite{PhysRevB.53.10921,Lifshitz1993,tarasenko98} and impact ionization of excitons~\cite{Ashkinadze:1988aa}.
%The inevitable heating of the electron gas leads to the excitation of processes associated with the decay of the trion state due to the Coulomb interaction between the high-energy electron and the trion. 
The matrix element of this process can be written as follows:
\begin{multline}
\label{Mub}
    M_{ub\pm} = \frac{4\pi^{3/2}\kappa}{S^{3/2}}\left[V_c(|\bm k_1 - \bm k|) \pm V_c(|\bm k_2 - \bm k|)\right]\\ \times \frac{\delta_{\bm k_1 + \bm k_2 + \bm k_x, \bm k + \bm K}}{\kappa^2+\left(\bm k_x - \frac{m_x}{m_x + m_e}\bm K\right)^2}.
\end{multline}
Here $V_c(q)$ is the matrix element of the Coulomb interaction, $\bm k$ is the initial wavevector of the incident (high-energy) electron, $\bm K$ is the wavevector of the trion (Fermi polaron), $\bm k_x$ is the wavevector of the exciton in the final state, $\bm k_1, \bm k_2$ are the final wavevectors of the initially bound and free electron, $\kappa = \sqrt{2\mu E_T/\hbar^2}$. In Eq.~\eqref{Mub} the $\pm$ signs correspond to the singlet and triplet states of a free electron and an electron bound to an exciton described, respectively, by the symmetric and antisymmetric wavefunctions. As before, the internal structure of exciton is disregarded. In this part,  we do not take into account Fermi-polaron effects (see Appendix~\ref{app:details}) and we disregard the interaction-induced modification of the free state wavefunctions taking them as plane waves. It is possible  to neglect here the Fermi-polaron correlations because the transition probability is exponentially small for small wavevectors and only high-momentum states are involved with $k\sim \kappa \gg k_F$. We also note that for non-degenerate electron gas the the role of Fermi sea diminishes~\cite{PhysRevB.108.125406}.   

\begin{figure}[t]
    \centering
    \includegraphics[width=\linewidth]{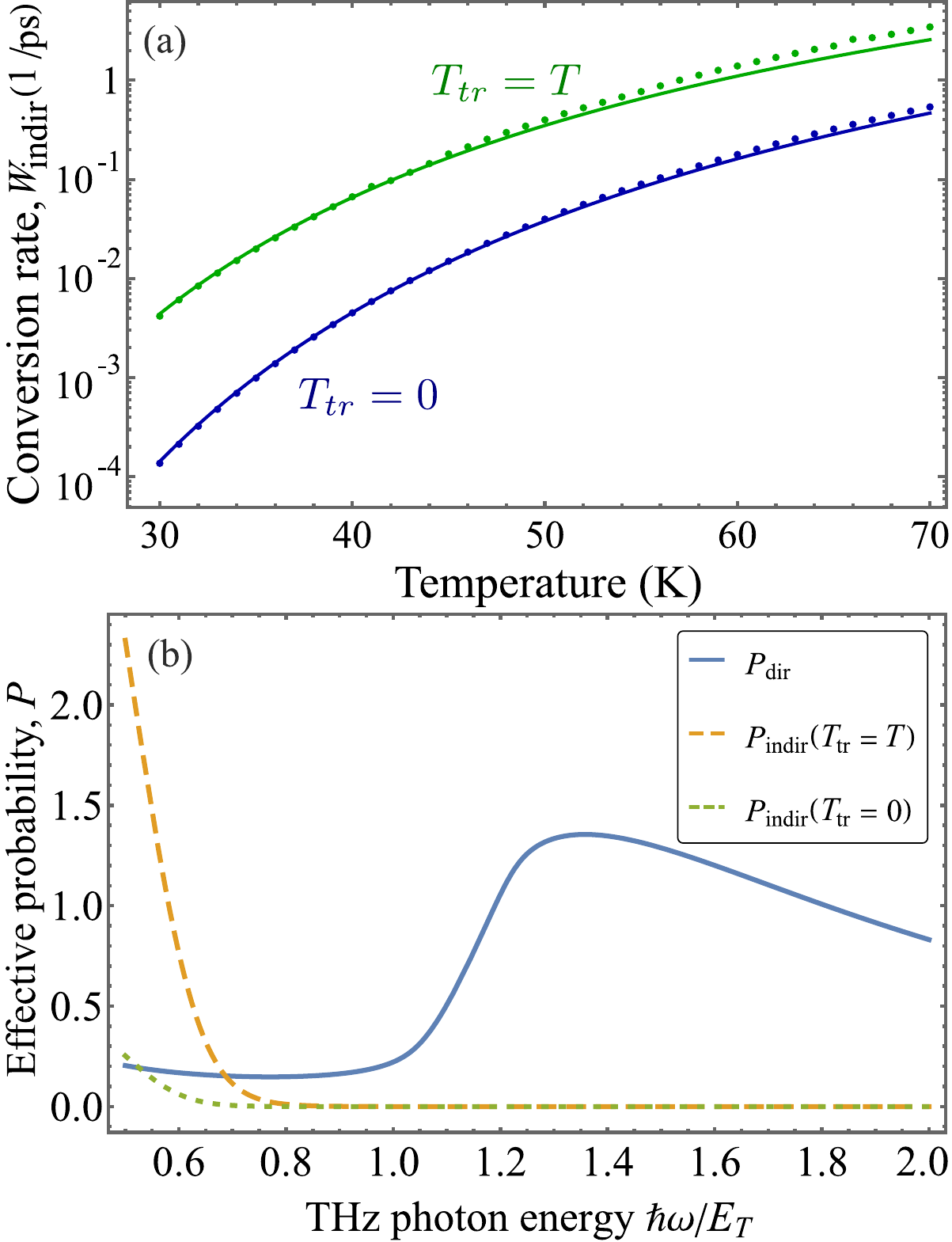}
    \caption{(a) Dependence of collision-induced transition rate $W_{\rm indir}$, Eq.~\eqref{W:ub}, on the electron gas temperature. Blue dots correspond to the transition rate for a trion with $\bm K = 0$, green dots correspond the same temperatures of electrons and trions. Solid lines are plotted after the approximate Eq.~\eqref{W:ub:approx} with electron density $n_e = 5.8\times10^{11}$~cm$^{-2}$ ($E_F = 0.1E_T$) and prefactors $w_0 = 1.16\times 10^3$~cm$^{2}$s$^{-1}$ and $0.76\times 10^3$~cm$^{2}$s$^{-1}$ found from the best fit of the numerical calculation. (b) Dependence of effective probability $P$~\eqref{P} on the THz radiation frequency, with fluence $\phi = 5$~$\mu$J/cm$^{2}$ and $\tau_{\rm THZ} = 4$~ps. The dependence of electron temperature $T(\omega)$ for $P_{indir}(T)$ corresponds to Fig.~\ref{fig2} with $\tau = 2$ ps.}
    \label{fig3}
\end{figure}

We consider the electron gas described by the distribution function $f_{\bm k}$ that corresponds to a certain temperature $T$ and the non-degenerate gas of trions at the temperature $T_{tr}$ described by the Boltzmann distribution $f_{\bm K}^{T_{tr}}$. The trion density is $n_T$. Thus, using Fermi's Golden Rule we obtain the trion to exciton conversion rate caused by the trion-electron transitions
\begin{multline}
\label{W:ub}
    W_{\rm indir} = \frac{2\pi}{\hbar} \sum_{\substack{\bm k, \bm K\\ {\bm k_x, \bm k_1,\bm k_2}}} \left(\frac{3}{4}|M_{ub-}|^2+\frac{1}{4}|M_{ub+}|^2\right)\frac{f^{T_{tr}}_{\bm K}}{n_T}f_{\bm k}\\
    \delta(\varepsilon_{\bm k}+\varepsilon^T_{\bm K} -\varepsilon^X_{\bm k_x}-\varepsilon_{\bm k_1}-\varepsilon_{\bm k_2} - E_{T}).
\end{multline}
In our numerical calculations, we used the Coulomb potential 
\begin{equation}
\label{V:Cl}
    V_c(q) = \frac{e^2}{\epsilon q},
\end{equation}
where $\epsilon$ is the is the effective dielectric constant (in Fig.~\ref{fig3} $\epsilon = 4$). Note that since typical involved wavevectors $q\sim \varkappa$ (the inverse trion Bohr radius) the screening effects are not particularly important.

Figure~\ref{fig3} shows a strong exponential dependence of $W_{\rm indir}$ on the electron gas temperature $T$. It is because for the relevant temperature range only the quasiparticles from the high-energy tails of the distribution functions have sufficient energy to participate in a collision accompanied by the trion disintegration. Disregarding the wavevector dependence of the matrix elements~\eqref{Mub} we derive the following expression for the conversion rate
\begin{equation}
    \label{W:ub:approx}
    W_{\rm indir} \approx w_0n_e \frac{k_B T}{E_T} \exp{\left(- \frac{\alpha E_T}{k_B T} \right)},
\end{equation}
where the dimensional prefactor $w_0$ and dimensionless factor $\alpha$ in the exponent weakly depend on the trion and electron gas temperatures. Particularly, for equal electron and trion temperatures $\alpha=1$ and for cooled trions ($T_{tr}=0$ or, equivalently, we consider trion with $\bm K=0$) $\alpha = 1+ M_e/M_T$. The quantity $\alpha E_T$ has the meaning of the effective reaction threshold. Such exponential dependence is typical for the impact ionization processes. Analytical expression~\eqref{W:ub:approx} describes the transition rate $W_{\rm indir}$ quite well and stops working at temperatures $k_BT{\to} E_T$ where exponent saturates, see slight deviations in Fig.~\ref{fig3}(a) for highest temperatures. Importantly, at temperatures $T\gtrsim 50$~K the transition rate $W_{\rm indir}$ begins to be significant compared to the direct THz-induced transition rate from the trion (attractive polaron) to the exciton (repulsive polaron) $W_{\rm dir}$ shown in Fig.~\ref{fig1}.
To compare the efficiency of direct and indirect processes, we propose to consider the following integral value:
\begin{equation}
\label{P}
    P = \int_{0}^{\infty} W(t)dt,
\end{equation}
where $W$ is the rate of a direct ($W\equiv W_{\rm dir}$) or indirect ($W\equiv W_{\rm indir}$) process.  For small $W$, the quantity $P$ gives the probability of an attractive polaron to be converted to a repulsive one per pulse. The values $P>1$ correspond to the need of accounting for saturation processes and more detailed kinetics of polarons. We abstain from simulation of the full dynamics of attractive and repulsive polarons including nonlinearities that needs also to account for the polaron formation rates~\cite{PhysRevB.102.125410,Glazov:2020wf} and use $P$ as a simple yet physically justified measure of relative efficiency of the processes. In Eq.~\eqref{P}, the transition rate $W_{\rm dir}$ is a constant only during the pulse action and $P_{\rm dir} = W_{\rm dir}\tau_{\rm THz}$. For the indirect process, $W_{\rm indir}$ depends on time via the electron temperature which is found from Eq.~\eqref{heat:balance}, see Fig.~\ref{fig2}(b). Note that the time dependence of $W_{\rm indir}$ shown in Fig.~\ref{fig2}(b) at short times is delayed compared to the temperature dependence. It is because $W_{\rm indir}$ depends exponentially on the temperature. For the same reason, as soon as the THz pulse is over, the $W_{\rm indir}$ drops significantly faster than the temperature. The dependence of the effective probability is shown in Fig.~\eqref{fig3}(b). As expected, $P_{\rm dir}$ shows a resonant behavior repeating the THz photon energy dependence of $W_{\rm dir}$ in Fig.~\ref{fig1}. By contrast, $P_{\rm indir}$ has a sharp peak at $\omega\to 0$ where the electron heating up to $T \approx 50-60$~K is possible, see Fig.~\ref{fig2}.

%we also perturbatively took into account correlations with the Fermi sea in the form of a term in energy and summation with some function $C(\bm q)$ such that

%\begin{equation}
%    \sum_{\bm q} |C(\bm q)|^2 = 1,
%\end{equation}
%this may be important at temperatures low compared to the Fermi energy, but becomes less significant as we approach Boltzmann statistics. Thus, for zero temperature in Eq.~\eqref{M:TX},~\eqref{W:TX} this place was occupied by the following expression:
%\begin{equation}
%   |C_0(\bm q)|^2 = \frac{\frac{\pi\hbar^2M_X^2E_F}{2SM_T^2M_e}\sinh^{-2}\left(\frac{1}{2}\frac{M^2_X}{M^2_T}\right)\theta(E_F- \varepsilon_{\bm q})}{(E_T + E_{FP}+\varepsilon_{\bm q_0}-\varepsilon_{\bm q_0}^T-\frac{M_T}{M_x}E_F)^2}.
%\end{equation}
%For neglecting correlations, Eq.~\eqref{W:ub} would have to be summed with $|C_n(\bm q)|^2 = \delta_{\bm q,0}$, we also propose to consider the Poisson distribution $|C_n(\bm q)|^2 = 2\pi \hbar^2\theta(E_F-\varepsilon_{\bm q})/M_eE_FS$ as an example, see Fig.3(a). Decay due to high-energy electrons becomes important at temperatures $ T> 40$ K and can account for tenths of the process with light absorption.

\section{Conclusion}\label{sec:concl}

In this work, we have developed a theoretical framework to describe terahertz (THz) radiation-induced conversion between attractive and repulsive Fermi polaron states—corresponding to trions and excitons—in transition metal dichalcogenide monolayers. Our analysis goes beyond the basic trion picture and incorporates many-body correlations with the Fermi sea of resident charge carriers.

We show that the direct THz absorption process leading to polaron conversion exhibits a characteristic frequency dependence near the threshold, with a $(\hbar\omega - |E_{FP}|)^{3/2}$ scaling arising from the exciton correlations with the Fermi sea. At higher frequencies, the conversion rate aligns with the trion-based model. We also account for the effect of spectral broadening due to disorder and phonon scattering, which smoothens the absorption threshold.

Furthermore, we demonstrate that intense THz pulses can significantly heat the electron gas via Drude absorption, and this heating gives rise to an additional conversion mechanism via collisions between high-energy electrons and polarons. This indirect process exhibits a strong exponential dependence on temperature and becomes comparable to the direct optical conversion at electron temperatures above $\sim 50$~K which can be achieved at low THz photon energies.

Our results highlight the importance of many-body correlations and thermal effects in interpreting THz-induced exciton-trion dynamics and provide quantitative predictions for future experiments aimed at controlling excitonic states in two-dimensional semiconductors including emerging systems of van der Waals magnets such as CrSBr where the trion binding energies also correspond to the THz spectral range~\cite{Tabataba-Vakili:2024aa,PhysRevB.111.205301}. 

\acknowledgments

The authors are grateful to A. Chernikov, M.~Cuccu, E. Malic, and Z. Iakovlev for valuable discussions. This work was supported by RSF Grant No. 23-12-00142 (M.M.G., analytical model). A.M.S. acknowledges support of the Ministry of Science and Higher Education of the Russian Federation, Project No. FFWR-2024-0017 (numerical calculations).

%\newpage 
\appendix
\section{Technical details}\label{app:details}

\subsection{Fermi polaron wavefunctions and diagrams}

\begin{figure}[b]
    \centering
    \includegraphics[width=0.85\linewidth]{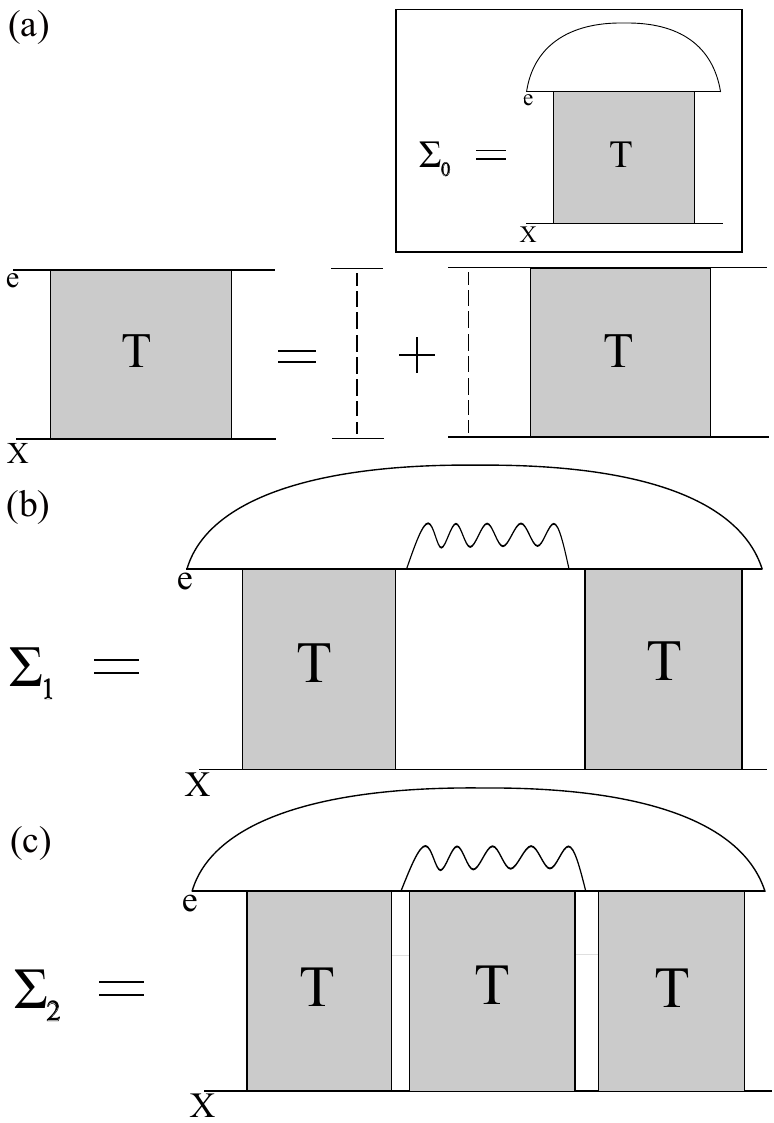}
    \caption{(a) The diagrammatic series defining the $T$ matrix in the Fermi polaron approach~\cite{cotlect2019transport,kadanoff2018quantum,PhysRevB.105.075311}. The lower line $X$ on Fig.~\ref{fig4} denotes the bare Green's function $G^0_X$ of the exciton, the upper $e$ is the bare Green's function of the electron $G^0_e$. The dotted line is the electron-exciton interaction $V$. The insert is the self-energy part $\Sigma_0$ defining the energy of the Fermi polaron. (b) $\Sigma_1$ is the contribution in self-energy determining the asymptotics of absorption spectra at $\hbar\omega - |E_{FP}| \gg E_F$, which does not take into account the Fermi-polar corrections in the matrix element Eq.~\eqref{M:TX}. The wavy line represents the interaction of the electron with the classical field~\eqref{Hle}. (c) $\Sigma_2$ is the Fermi-polaron correction to $\Sigma_1$.}
    \label{fig4}
\end{figure}

Expressions for $\varphi_{\bm k}, F_{\bm k}(\bm p, \bm q)$ in Eq.~\eqref{Psi:k} and their derivation based on the Schr\"odinger equation can be found in  Refs.~\cite{iakovlev2023fermi,iakovlev2024longitudinal}. In the relevant limit $\bm k \to 0$ they are written as:
\begin{equation}
    \label{varphi}
    \varphi_0^2 = \frac{1}{4\frac{E_T}{E_F}\left(\frac{M_T}{M_X}\right)^3\sinh^2{\left[\frac{1}{2}\left(\frac{M_X}{M_T}\right)^2\right]}},
\end{equation}
\begin{multline}
    \label{F_0}
    F_{0}(\bm p, \bm q) = \frac{2\pi\varphi_0{\hbar^2}E_T}{\mu S}
    \\
    \times\frac{(E_T + E_{FP}+\varepsilon_{\bm q}-\varepsilon_{\bm q}^T-\frac{M_T}{M_X}E_F)^{-1}}{(E_{FP}-\varepsilon^X_{-\bm p + \bm q}-\varepsilon_{\bm p} + \varepsilon_{\bm q})}.
\end{multline}

To find $U_{\bm k = 0, \bm p, \bm q}(\bm p')$ from Eq.\eqref{Phi:k} for the repulsive polaron, we use similar principles as for finding the coefficients for the attractive polaron:
\begin{multline}
    \label{U}
    U_{\bm k = 0, \bm p, \bm q}(\bm p') = \frac{\xi^{-1}({\bm p, \bm q})}{S\left(\varepsilon_{-\bm p+\bm q}^{X}+ \varepsilon_{\bm p} -\varepsilon_{-\bm p'+\bm q}^{X} - \varepsilon_{\bm p'} \right)},
\end{multline}
with
\begin{multline}
    \label{xi}
    \xi({\bm p, \bm q}) = -\frac{\mu }{2\pi{\hbar^2}}\ln{(E_{X}/E_{T})} \\- \frac{1}{S}\sum_{\bm p'} \frac{1}{\varepsilon_{-\bm p+\bm q}^{X} + \varepsilon_{\bm p}  -\varepsilon_{-\bm p'+\bm q}^{X} - \varepsilon_{\bm p'}},
\end{multline}
where $E_X$ is exciton binding energy, the sum over $\bm p'$ is limited by the cutoff parameter and corresponds to the inverse size of the exciton, and we use approximation~\cite{suris:correlation} for parameter exciton-electron interaction $V = {-}\frac{2\pi{\hbar^2}}{ \mu S }\ln{{^{-1}}(E_{X}/E_{T})}$. Despite the fact that $U_{\bm k, \bm p, \bm q}(\bm p') \sim 1/\ln{(E_T/E_F)}$ for small $p\sim\bm k_F$, this smallness disappears when summing over the electron momentum $\bm p'$ in Eq.~\eqref{me}.  Using the expressions~(\ref{varphi}-\ref{xi}) we obtain the expression for the matrix element Eq.~\eqref{M:TX}.

Same results can be derived diagrammatically, see Fig.~\ref{fig4}.  The main contribution determining the absorption spectrum to the self-energy part is given by the following diagrams Fig.~\ref{fig4}(b,c).

The Green function of an exciton near the pole corresponding to the attractive Fermi polaron can be represented as follows:
\begin{equation}
    G_X(E,\bm k) = \frac{Z}{E - E_{\bm k}  + iW_{dir}/2+ i\Gamma}.
\end{equation}
Here $E_{\bm k}$ is the Fermi polaron energy with momentum $\bm k$ and the weight of the pole $Z$ has the form
\begin{equation}
    Z = \left(1 - \frac{\partial \text{Re}\Sigma(E, \bm k)}{\partial E}\bigg |_{E = E_{\bm k}}\right)^{-1},
\end{equation}
where $\Sigma$ is total self-energy, in particularly $\Sigma = \Sigma_0 + \Sigma_1 + \Sigma_2$.
The transition frequency can be represented as
\begin{equation}
    W_{\rm dir} = -I\times 2Z\frac{\partial \text{Im} \Sigma(E_{\bm k}, \bm k)}{\partial I}\bigg|_{I \to 0}.
\end{equation}

\newpage 
\subsection{Evaluation of $W_{\rm dir}$ and limit $E_F \to 0$}

For Eq.~\eqref{W:TX} the calculation can be reduced to a one-dimensional integral:
\begin{widetext}
\begin{multline}
    W_{\rm dir}(\omega) = I\times\frac{\alpha\pi\mu\hbar^2 E_TE_F}{M_e^2(\hbar\omega)^4\sinh^2\left[\frac{1}{2}\left(\frac{M_X}{M_T}\right)^2\right]}\int_{0}^{E_F} du\frac{\frac{M_X}{M_T}u+\hbar\omega+E_{FP}}{\left(\frac{E_F}{1-e^{-\left(\frac{M_X}{M_T}\right)^2}}-u\right)^2}
    \mbox{Re}\Biggl[\arccos\left({\frac{E_F-\frac{M_X-M_e}{M_T}u-\frac{M_X}{M_T}(\hbar\omega+E_{FP})}{2\frac{M_e}{M_T}\sqrt{E_Fu}}}\right)\\+
    \arctan\Biggl(\frac{\sqrt{(2\frac{M_e}{M_T}\sqrt{uE_F})^2-[E_F-\frac{M_X-M_e}{M_T}u-\frac{M_X}{M_T}(\hbar\omega+E_{FP})]^2}}{E_F+\frac{M_X-M_e}{M_T}u+\frac{M_X}{M_T}(\hbar\omega+E_{FP})}\Biggr)\Biggr].
\end{multline}
\end{widetext}
Here we introduced inverse trigonometric functions in the complex plane. In the case $\hbar\omega + E_{FP} > 4M_eE_F/M_X$ the real part of the expression in brackets is $\pi$, and $W_{\rm dir}$ takes the form Eq.~\eqref{W:TP}. Besides this, in Eq.~\eqref{W:TP} $| E_{FP}+\beta E_F + E_T|/E_F {\approx \frac{1}{12}(M_X/M_T)^3 + O((M_X/M_T)^7/2^{7/2})} \ll 1$ for existing $M_X/M_T$. For the case shown in Fig.~\ref{fig1}(a) this contribution equals $0.03$. Thus in the limit $\ln(E_T/E_E) \gg 1$ the solution has a weak dependence on $E_F$ on the same scale. This result can be obtained from the problem of two bodies with short-range interaction:
\begin{equation}
    \left(-\frac{\nabla^2_{\bm R}}{2M_T} - \frac{\nabla^2_{\bm r}}{2\mu} + V(\bm r)\right)\psi(\bm R,\bm r) = E\psi(\bm R,\bm r),
\end{equation}
where $\bm R = (M_e \bm r_e + M_X \bm r_X)/M_T$ is center of mass position-vector, $\bm r = \bm r_e - \bm r_X$ is position-vector of relative motion and $V(\bm r)$ is short-range interaction between electron and exciton, corresponding to the shallow well problem~\cite{landau2013quantum, levitov2003green}. The wave function $\psi(\bm R, \bm r) = e^{i\bm P\bm R}\psi(\bm r)$, where $\bm P$ is the total moment of the system. The relative motion part of wave function for a bound state is~\cite{Glazov:2020wf}
\begin{equation}
    \psi_{b}(\bm r) = \frac{\kappa}{\sqrt{\pi}}K_0(\kappa r).
\end{equation}
Here $\kappa = \sqrt{2\mu E_T}/\hbar$ corresponds to the inverse Bohr radius of the trion, $K_0(\kappa r)$ is the Macdonald function. For unbound (scattering) states the wave function has the form~\cite{Glazov:2020wf}
\begin{equation}
    \psi_{\bm k}(\bm r) = \frac{1}{\sqrt{S}}\left(e^{i\bm k \bm r} -\frac{i\mu}{2}S\left(\frac{\hbar^2k^2}{2\mu}\right)H_0^{(1)}(kr)\right),
\end{equation}
with
\begin{equation}
    S\left(\varepsilon\right) = -\frac{2\pi}{\mu\ln{\left(\frac{\varepsilon+i\delta}{-E_{T}}\right)}},
\end{equation}
where $k$ is the relative motion wavevector, $H_0^{(1)}(kr)$ is the Hankel function of the first kind and $S\left(\varepsilon\right)$ is the scattering amplitude in the $s$-channel.
If we ignore scattering in the $p$-channel, then the matrix element for ground bound state is proportional to the Fourier transform $\psi_b(\bm r)$, in particular:
\begin{equation}
    M_{\rm dir} = {-}\sqrt{\frac{\kappa^2}{\pi}}(\bm k \cdot \bm A)\frac{e{\hbar}}{\sqrt{S}cM_e}\int K_0(\kappa r) e^{i\bm k \bm r} d^2r
\end{equation}
The transition rate $W_{\rm dir}$ corresponds to the limit $E_F \to 0 $ in Eq.~\eqref{W:TP} and is equal to
\begin{equation}
     W_{\rm dir} = I\times  \frac{4\alpha \pi^2\mu\hbar^2 E_{T}}{M_e^2}
    \frac{\hbar\omega - E_{T}}{(\hbar\omega)^4}\theta(\hbar\omega - E_{T}).
\end{equation}

Our approach leads to a Bessel-type wave function. If we assume that the decay of the wave function for the bound state is purely exponential $\psi_b(\bm r)\sim e^{-\kappa r}$, then the power-law dependence on $\omega$ for the transition frequency changes:
\begin{equation}
\label{Wdir:exp}
    W^e_{\rm dir} \sim \frac{\hbar\omega - E_T}{(\hbar\omega)^5}.
\end{equation}
In the general case, $W_{\rm dir} \sim (\hbar\omega - E_T)/g(\hbar\omega, E_T)$, where $g$ strongly depends on the form of the wave function of the bound state at $r \sim 1/\kappa$.

\newpage
%\nocite{*}
\let\itshape\upshape 
%\bibliography{references}% Produces the bibliography via BibTeX.
%\bibliography{all-1}% Produces the bibliography via BibTeX.

\begin{thebibliography}{54}%
\makeatletter
\providecommand \@ifxundefined [1]{%
 \@ifx{#1\undefined}
}%
\providecommand \@ifnum [1]{%
 \ifnum #1\expandafter \@firstoftwo
 \else \expandafter \@secondoftwo
 \fi
}%
\providecommand \@ifx [1]{%
 \ifx #1\expandafter \@firstoftwo
 \else \expandafter \@secondoftwo
 \fi
}%
\providecommand \natexlab [1]{#1}%
\providecommand \enquote  [1]{``#1''}%
\providecommand \bibnamefont  [1]{#1}%
\providecommand \bibfnamefont [1]{#1}%
\providecommand \citenamefont [1]{#1}%
\providecommand \href@noop [0]{\@secondoftwo}%
\providecommand \href [0]{\begingroup \@sanitize@url \@href}%
\providecommand \@href[1]{\@@startlink{#1}\@@href}%
\providecommand \@@href[1]{\endgroup#1\@@endlink}%
\providecommand \@sanitize@url [0]{\catcode `\\12\catcode `\$12\catcode
  `\&12\catcode `\#12\catcode `\^12\catcode `\_12\catcode `\%12\relax}%
\providecommand \@@startlink[1]{}%
\providecommand \@@endlink[0]{}%
\providecommand \url  [0]{\begingroup\@sanitize@url \@url }%
\providecommand \@url [1]{\endgroup\@href {#1}{\urlprefix }}%
\providecommand \urlprefix  [0]{URL }%
\providecommand \Eprint [0]{\href }%
\providecommand \doibase [0]{https://doi.org/}%
\providecommand \selectlanguage [0]{\@gobble}%
\providecommand \bibinfo  [0]{\@secondoftwo}%
\providecommand \bibfield  [0]{\@secondoftwo}%
\providecommand \translation [1]{[#1]}%
\providecommand \BibitemOpen [0]{}%
\providecommand \bibitemStop [0]{}%
\providecommand \bibitemNoStop [0]{.\EOS\space}%
\providecommand \EOS [0]{\spacefactor3000\relax}%
\providecommand \BibitemShut  [1]{\csname bibitem#1\endcsname}%
\let\auto@bib@innerbib\@empty
%</preamble>
\bibitem [{\citenamefont {Ivchenko}(2005)}]{ivchenko2005optical}%
  \BibitemOpen
  \bibfield  {author} {\bibinfo {author} {\bibfnamefont {E.~L.}\ \bibnamefont
  {Ivchenko}},\ }\href@noop {} {\emph {\bibinfo {title} {Optical spectroscopy
  of semiconductor nanostructures}}}\ (\bibinfo  {publisher} {Alpha Science,
  Harrow},\ \bibinfo {year} {2005})\BibitemShut {NoStop}%
\bibitem [{\citenamefont {Haug}\ and\ \citenamefont
  {Koch}(2009)}]{haug2009quantum}%
  \BibitemOpen
  \bibfield  {author} {\bibinfo {author} {\bibfnamefont {H.}~\bibnamefont
  {Haug}}\ and\ \bibinfo {author} {\bibfnamefont {S.~W.}\ \bibnamefont
  {Koch}},\ }\href@noop {} {\emph {\bibinfo {title} {Quantum theory of the
  optical and electronic properties of semiconductors}}}\ (\bibinfo
  {publisher} {World Scientific},\ \bibinfo {year} {2009})\BibitemShut
  {NoStop}%
\bibitem [{\citenamefont {Klingshirn}(2012)}]{klingshirn2012semiconductor}%
  \BibitemOpen
  \bibfield  {author} {\bibinfo {author} {\bibfnamefont {C.~F.}\ \bibnamefont
  {Klingshirn}},\ }\href@noop {} {\emph {\bibinfo {title} {Semiconductor
  optics}}}\ (\bibinfo  {publisher} {Springer Science \& Business Media,
  Heidelberg},\ \bibinfo {year} {2012})\BibitemShut {NoStop}%
\bibitem [{\citenamefont {Semina}\ and\ \citenamefont
  {Suris}(2022)}]{Semina_2022}%
  \BibitemOpen
  \bibfield  {author} {\bibinfo {author} {\bibfnamefont {M.~A.}\ \bibnamefont
  {Semina}}\ and\ \bibinfo {author} {\bibfnamefont {R.~A.}\ \bibnamefont
  {Suris}},\ }\bibfield  {title} {\bibinfo {title} {Localized excitons and
  trions in semiconductor nanosystems},\ }\href
  {https://doi.org/10.3367/ufne.2020.11.038867} {\bibfield  {journal} {\bibinfo
   {journal} {Physics-Uspekhi}\ }\textbf {\bibinfo {volume} {65}},\ \bibinfo
  {pages} {111} (\bibinfo {year} {2022})}\BibitemShut {NoStop}%
\bibitem [{\citenamefont {Durnev}\ and\ \citenamefont
  {Glazov}(2018)}]{Durnev:2018}%
  \BibitemOpen
  \bibfield  {author} {\bibinfo {author} {\bibfnamefont {M.~V.}\ \bibnamefont
  {Durnev}}\ and\ \bibinfo {author} {\bibfnamefont {M.~M.}\ \bibnamefont
  {Glazov}},\ }\bibfield  {title} {\bibinfo {title} {Excitons and trions in
  two-dimensional semiconductors based on transition metal dichalcogenides},\
  }\href {https://doi.org/10.3367/UFNe.2017.07.038172} {\bibfield  {journal}
  {\bibinfo  {journal} {Phys. Usp.}\ }\textbf {\bibinfo {volume} {61}},\
  \bibinfo {pages} {825вЂ“845} (\bibinfo {year} {2018})}\BibitemShut {NoStop}%
\bibitem [{\citenamefont {Wang}\ \emph {et~al.}(2018)\citenamefont {Wang},
  \citenamefont {Chernikov}, \citenamefont {Glazov}, \citenamefont {Heinz},
  \citenamefont {Marie}, \citenamefont {Amand},\ and\ \citenamefont
  {Urbaszek}}]{RevModPhys.90.021001}%
  \BibitemOpen
  \bibfield  {author} {\bibinfo {author} {\bibfnamefont {G.}~\bibnamefont
  {Wang}}, \bibinfo {author} {\bibfnamefont {A.}~\bibnamefont {Chernikov}},
  \bibinfo {author} {\bibfnamefont {M.~M.}\ \bibnamefont {Glazov}}, \bibinfo
  {author} {\bibfnamefont {T.~F.}\ \bibnamefont {Heinz}}, \bibinfo {author}
  {\bibfnamefont {X.}~\bibnamefont {Marie}}, \bibinfo {author} {\bibfnamefont
  {T.}~\bibnamefont {Amand}},\ and\ \bibinfo {author} {\bibfnamefont
  {B.}~\bibnamefont {Urbaszek}},\ }\bibfield  {title} {\bibinfo {title}
  {Colloquium: Excitons in atomically thin transition metal dichalcogenides},\
  }\href {https://doi.org/10.1103/RevModPhys.90.021001} {\bibfield  {journal}
  {\bibinfo  {journal} {Rev. Mod. Phys.}\ }\textbf {\bibinfo {volume} {90}},\
  \bibinfo {pages} {021001} (\bibinfo {year} {2018})}\BibitemShut {NoStop}%
%
\bibitem{bruun}
P.~Massignan, R.~Schmidt, G.~E. Astrakharchik, A.~Imamoglu, M.~Zwierlein,
  J.~J. Arlt, and G.~M. Bruun, Polarons in atomic gases and two-dimensional semiconductors, \href{https://arxiv.org/abs/2501.09618}{arXiv:2501.09618} (2025).
%
\bibitem [{\citenamefont {Mak}\ \emph {et~al.}(2013)\citenamefont {Mak},
  \citenamefont {He}, \citenamefont {Lee}, \citenamefont {Lee}, \citenamefont
  {Hone}, \citenamefont {Heinz},\ and\ \citenamefont {Shan}}]{Mak:2013lh}%
  \BibitemOpen
  \bibfield  {author} {\bibinfo {author} {\bibfnamefont {K.~F.}\ \bibnamefont
  {Mak}}, \bibinfo {author} {\bibfnamefont {K.}~\bibnamefont {He}}, \bibinfo
  {author} {\bibfnamefont {C.}~\bibnamefont {Lee}}, \bibinfo {author}
  {\bibfnamefont {G.~H.}\ \bibnamefont {Lee}}, \bibinfo {author} {\bibfnamefont
  {J.}~\bibnamefont {Hone}}, \bibinfo {author} {\bibfnamefont {T.~F.}\
  \bibnamefont {Heinz}},\ and\ \bibinfo {author} {\bibfnamefont
  {J.}~\bibnamefont {Shan}},\ }\bibfield  {title} {\bibinfo {title} {Tightly
  bound trions in monolayer {MoS}$_2$},\ }\href
  {http://dx.doi.org/10.1038/nmat3505} {\bibfield  {journal} {\bibinfo
  {journal} {Nat Mater}\ }\textbf {\bibinfo {volume} {12}},\ \bibinfo {pages}
  {207} (\bibinfo {year} {2013})}\BibitemShut {NoStop}%
\bibitem [{\citenamefont {Courtade}\ \emph {et~al.}(2017)\citenamefont
  {Courtade}, \citenamefont {Semina}, \citenamefont {Manca}, \citenamefont
  {Glazov}, \citenamefont {Robert}, \citenamefont {Cadiz}, \citenamefont
  {Wang}, \citenamefont {Taniguchi}, \citenamefont {Watanabe}, \citenamefont
  {Pierre}, \citenamefont {Escoffier}, \citenamefont {Ivchenko}, \citenamefont
  {Renucci}, \citenamefont {Marie}, \citenamefont {Amand},\ and\ \citenamefont
  {Urbaszek}}]{Courtade:2017a}%
  \BibitemOpen
  \bibfield  {author} {\bibinfo {author} {\bibfnamefont {E.}~\bibnamefont
  {Courtade}}, \bibinfo {author} {\bibfnamefont {M.}~\bibnamefont {Semina}},
  \bibinfo {author} {\bibfnamefont {M.}~\bibnamefont {Manca}}, \bibinfo
  {author} {\bibfnamefont {M.~M.}\ \bibnamefont {Glazov}}, \bibinfo {author}
  {\bibfnamefont {C.}~\bibnamefont {Robert}}, \bibinfo {author} {\bibfnamefont
  {F.}~\bibnamefont {Cadiz}}, \bibinfo {author} {\bibfnamefont
  {G.}~\bibnamefont {Wang}}, \bibinfo {author} {\bibfnamefont {T.}~\bibnamefont
  {Taniguchi}}, \bibinfo {author} {\bibfnamefont {K.}~\bibnamefont {Watanabe}},
  \bibinfo {author} {\bibfnamefont {M.}~\bibnamefont {Pierre}}, \bibinfo
  {author} {\bibfnamefont {W.}~\bibnamefont {Escoffier}}, \bibinfo {author}
  {\bibfnamefont {E.~L.}\ \bibnamefont {Ivchenko}}, \bibinfo {author}
  {\bibfnamefont {P.}~\bibnamefont {Renucci}}, \bibinfo {author} {\bibfnamefont
  {X.}~\bibnamefont {Marie}}, \bibinfo {author} {\bibfnamefont
  {T.}~\bibnamefont {Amand}},\ and\ \bibinfo {author} {\bibfnamefont
  {B.}~\bibnamefont {Urbaszek}},\ }\bibfield  {title} {\bibinfo {title}
  {Charged excitons in monolayer {WSe}$_{2}$: Experiment and theory},\ }\href
  {https://doi.org/10.1103/PhysRevB.96.085302} {\bibfield  {journal} {\bibinfo
  {journal} {Phys. Rev. B}\ }\textbf {\bibinfo {volume} {96}},\ \bibinfo
  {pages} {085302} (\bibinfo {year} {2017})}\BibitemShut {NoStop}%
\bibitem [{\citenamefont {Arora}\ \emph {et~al.}(2019)\citenamefont {Arora},
  \citenamefont {Deilmann}, \citenamefont {Reichenauer}, \citenamefont {Kern},
  \citenamefont {Michaelis~de Vasconcellos}, \citenamefont {Rohlfing},\ and\
  \citenamefont {Bratschitsch}}]{PhysRevLett.123.167401}%
  \BibitemOpen
  \bibfield  {author} {\bibinfo {author} {\bibfnamefont {A.}~\bibnamefont
  {Arora}}, \bibinfo {author} {\bibfnamefont {T.}~\bibnamefont {Deilmann}},
  \bibinfo {author} {\bibfnamefont {T.}~\bibnamefont {Reichenauer}}, \bibinfo
  {author} {\bibfnamefont {J.}~\bibnamefont {Kern}}, \bibinfo {author}
  {\bibfnamefont {S.}~\bibnamefont {Michaelis~de Vasconcellos}}, \bibinfo
  {author} {\bibfnamefont {M.}~\bibnamefont {Rohlfing}},\ and\ \bibinfo
  {author} {\bibfnamefont {R.}~\bibnamefont {Bratschitsch}},\ }\bibfield
  {title} {\bibinfo {title} {Excited-state trions in monolayer {WS}$_{2}$},\
  }\href {https://doi.org/10.1103/PhysRevLett.123.167401} {\bibfield  {journal}
  {\bibinfo  {journal} {Phys. Rev. Lett.}\ }\textbf {\bibinfo {volume} {123}},\
  \bibinfo {pages} {167401} (\bibinfo {year} {2019})}\BibitemShut {NoStop}%
\bibitem [{\citenamefont {Goldstein}\ \emph {et~al.}(2020)\citenamefont
  {Goldstein}, \citenamefont {Wu}, \citenamefont {Chen}, \citenamefont
  {Taniguchi}, \citenamefont {Watanabe}, \citenamefont {Varga},\ and\
  \citenamefont {Yan}}]{doi:10.1063/5.0013092}%
  \BibitemOpen
  \bibfield  {author} {\bibinfo {author} {\bibfnamefont {T.}~\bibnamefont
  {Goldstein}}, \bibinfo {author} {\bibfnamefont {Y.-C.}\ \bibnamefont {Wu}},
  \bibinfo {author} {\bibfnamefont {S.-Y.}\ \bibnamefont {Chen}}, \bibinfo
  {author} {\bibfnamefont {T.}~\bibnamefont {Taniguchi}}, \bibinfo {author}
  {\bibfnamefont {K.}~\bibnamefont {Watanabe}}, \bibinfo {author}
  {\bibfnamefont {K.}~\bibnamefont {Varga}},\ and\ \bibinfo {author}
  {\bibfnamefont {J.}~\bibnamefont {Yan}},\ }\bibfield  {title} {\bibinfo
  {title} {Ground and excited state exciton polarons in monolayer {MoSe$_2$}},\
  }\href {https://doi.org/10.1063/5.0013092} {\bibfield  {journal} {\bibinfo
  {journal} {The Journal of Chemical Physics}\ }\textbf {\bibinfo {volume}
  {153}},\ \bibinfo {pages} {071101} (\bibinfo {year} {2020})},\ \Eprint
  {https://arxiv.org/abs/https://doi.org/10.1063/5.0013092}
  {https://doi.org/10.1063/5.0013092} \BibitemShut {NoStop}%
\bibitem [{\citenamefont {Wagner}\ \emph {et~al.}(2020)\citenamefont {Wagner},
  \citenamefont {Wietek}, \citenamefont {Ziegler}, \citenamefont {Semina},
  \citenamefont {Taniguchi}, \citenamefont {Watanabe}, \citenamefont {Zipfel},
  \citenamefont {Glazov},\ and\ \citenamefont
  {Chernikov}}]{PhysRevLett.125.267401}%
  \BibitemOpen
  \bibfield  {author} {\bibinfo {author} {\bibfnamefont {K.}~\bibnamefont
  {Wagner}}, \bibinfo {author} {\bibfnamefont {E.}~\bibnamefont {Wietek}},
  \bibinfo {author} {\bibfnamefont {J.~D.}\ \bibnamefont {Ziegler}}, \bibinfo
  {author} {\bibfnamefont {M.~A.}\ \bibnamefont {Semina}}, \bibinfo {author}
  {\bibfnamefont {T.}~\bibnamefont {Taniguchi}}, \bibinfo {author}
  {\bibfnamefont {K.}~\bibnamefont {Watanabe}}, \bibinfo {author}
  {\bibfnamefont {J.}~\bibnamefont {Zipfel}}, \bibinfo {author} {\bibfnamefont
  {M.~M.}\ \bibnamefont {Glazov}},\ and\ \bibinfo {author} {\bibfnamefont
  {A.}~\bibnamefont {Chernikov}},\ }\bibfield  {title} {\bibinfo {title}
  {Autoionization and dressing of excited excitons by free carriers in
  monolayer {WSe$_{2}$}},\ }\href
  {https://doi.org/10.1103/PhysRevLett.125.267401} {\bibfield  {journal}
  {\bibinfo  {journal} {Phys. Rev. Lett.}\ }\textbf {\bibinfo {volume} {125}},\
  \bibinfo {pages} {267401} (\bibinfo {year} {2020})}\BibitemShut {NoStop}%
\bibitem [{\citenamefont {Lin}\ \emph {et~al.}(2022)\citenamefont {Lin},
  \citenamefont {Ziegler}, \citenamefont {Semina}, \citenamefont {Mamedov},
  \citenamefont {Watanabe}, \citenamefont {Taniguchi}, \citenamefont {Bange},
  \citenamefont {Chernikov}, \citenamefont {Glazov},\ and\ \citenamefont
  {Lupton}}]{Lin:2022aa}%
  \BibitemOpen
  \bibfield  {author} {\bibinfo {author} {\bibfnamefont {K.-Q.}\ \bibnamefont
  {Lin}}, \bibinfo {author} {\bibfnamefont {J.~D.}\ \bibnamefont {Ziegler}},
  \bibinfo {author} {\bibfnamefont {M.~A.}\ \bibnamefont {Semina}}, \bibinfo
  {author} {\bibfnamefont {J.~V.}\ \bibnamefont {Mamedov}}, \bibinfo {author}
  {\bibfnamefont {K.}~\bibnamefont {Watanabe}}, \bibinfo {author}
  {\bibfnamefont {T.}~\bibnamefont {Taniguchi}}, \bibinfo {author}
  {\bibfnamefont {S.}~\bibnamefont {Bange}}, \bibinfo {author} {\bibfnamefont
  {A.}~\bibnamefont {Chernikov}}, \bibinfo {author} {\bibfnamefont {M.~M.}\
  \bibnamefont {Glazov}},\ and\ \bibinfo {author} {\bibfnamefont {J.~M.}\
  \bibnamefont {Lupton}},\ }\bibfield  {title} {\bibinfo {title} {High-lying
  valley-polarized trions in {2D} semiconductors},\ }\href
  {https://doi.org/10.1038/s41467-022-33939-w} {\bibfield  {journal} {\bibinfo
  {journal} {Nature Communications}\ }\textbf {\bibinfo {volume} {13}},\
  \bibinfo {pages} {6980} (\bibinfo {year} {2022})}\BibitemShut {NoStop}%
\bibitem [{\citenamefont {Astakhov}\ \emph {et~al.}(2005)\citenamefont
  {Astakhov}, \citenamefont {Yakovlev}, \citenamefont {Rudenkov}, \citenamefont
  {Christianen}, \citenamefont {Barrick}, \citenamefont {Crooker},
  \citenamefont {Dzyubenko}, \citenamefont {Ossau}, \citenamefont {Maan},
  \citenamefont {Karczewski},\ and\ \citenamefont
  {Wojtowicz}}]{PhysRevB.71.201312}%
  \BibitemOpen
  \bibfield  {author} {\bibinfo {author} {\bibfnamefont {G.~V.}\ \bibnamefont
  {Astakhov}}, \bibinfo {author} {\bibfnamefont {D.~R.}\ \bibnamefont
  {Yakovlev}}, \bibinfo {author} {\bibfnamefont {V.~V.}\ \bibnamefont
  {Rudenkov}}, \bibinfo {author} {\bibfnamefont {P.~C.~M.}\ \bibnamefont
  {Christianen}}, \bibinfo {author} {\bibfnamefont {T.}~\bibnamefont
  {Barrick}}, \bibinfo {author} {\bibfnamefont {S.~A.}\ \bibnamefont
  {Crooker}}, \bibinfo {author} {\bibfnamefont {A.~B.}\ \bibnamefont
  {Dzyubenko}}, \bibinfo {author} {\bibfnamefont {W.}~\bibnamefont {Ossau}},
  \bibinfo {author} {\bibfnamefont {J.~C.}\ \bibnamefont {Maan}}, \bibinfo
  {author} {\bibfnamefont {G.}~\bibnamefont {Karczewski}},\ and\ \bibinfo
  {author} {\bibfnamefont {T.}~\bibnamefont {Wojtowicz}},\ }\bibfield  {title}
  {\bibinfo {title} {Definitive observation of the dark triplet ground state of
  charged excitons in high magnetic fields},\ }\href
  {https://doi.org/10.1103/PhysRevB.71.201312} {\bibfield  {journal} {\bibinfo
  {journal} {Phys. Rev. B}\ }\textbf {\bibinfo {volume} {71}},\ \bibinfo
  {pages} {201312} (\bibinfo {year} {2005})}\BibitemShut {NoStop}%
\bibitem [{\citenamefont {Jain}\ \emph {et~al.}(2025)\citenamefont {Jain},
  \citenamefont {Glazov},\ and\ \citenamefont {Arora}}]{8181-jzj5}%
  \BibitemOpen
  \bibfield  {author} {\bibinfo {author} {\bibfnamefont {S.}~\bibnamefont
  {Jain}}, \bibinfo {author} {\bibfnamefont {M.}~\bibnamefont {Glazov}},\ and\
  \bibinfo {author} {\bibfnamefont {A.}~\bibnamefont {Arora}},\ }\bibfield
  {title} {\bibinfo {title} {Excited-state trions in a quantum well},\ }\href
  {https://doi.org/10.1103/8181-jzj5} {\bibfield  {journal} {\bibinfo
  {journal} {Phys. Rev. Lett.}\ }\textbf {\bibinfo {volume} {134}},\ \bibinfo
  {pages} {246902} (\bibinfo {year} {2025})}\BibitemShut {NoStop}%
  %
  \bibitem{PhysRevLett.129.076801}
D.~Van~Tuan, S.-F. Shi, X.~Xu, S.~A. Crooker, and H.~Dery, Six-body and eight-body exciton states in monolayer
  ${\mathrm{WSe}}_{2}$, Phys. Rev. Lett. {\bf 129}, 076801 (2022).
%
\bibitem{ddn8-d8bs}
H.~Dery, C.~Robert, S.~A. Crooker, X.~Marie, and D.~Van~Tuan, Energy shifts and broadening of excitonic resonances in
  electrostatically doped semiconductors, Phys. Rev. X {\bf 15}, 031049 (2025).
%
\bibitem [{\citenamefont {Ganichev}\ \emph {et~al.}(2006)\citenamefont
  {Ganichev}, \citenamefont {Prettl},\ and\ \citenamefont
  {Prettl}}]{ganichev2006intense}%
  \BibitemOpen
  \bibfield  {author} {\bibinfo {author} {\bibfnamefont {S.}~\bibnamefont
  {Ganichev}}, \bibinfo {author} {\bibfnamefont {W.}~\bibnamefont {Prettl}},\
  and\ \bibinfo {author} {\bibfnamefont {W.}~\bibnamefont {Prettl}},\ }\href
  {https://books.google.de/books?id=95EUDAAAQBAJ} {\emph {\bibinfo {title}
  {Intense Terahertz Excitation of Semiconductors}}},\ Oxford science
  publications\ (\bibinfo  {publisher} {OUP Oxford},\ \bibinfo {year}
  {2006})\BibitemShut {NoStop}%
\bibitem [{\citenamefont {Lampin}\ \emph {et~al.}(2020)\citenamefont {Lampin},
  \citenamefont {Mouret}, \citenamefont {Dhillon},\ and\ \citenamefont
  {Mangeney}}]{Lampin2020}%
  \BibitemOpen
  \bibfield  {author} {\bibinfo {author} {\bibfnamefont {J.}~\bibnamefont
  {Lampin}}, \bibinfo {author} {\bibfnamefont {G.}~\bibnamefont {Mouret}},
  \bibinfo {author} {\bibfnamefont {S.}~\bibnamefont {Dhillon}},\ and\ \bibinfo
  {author} {\bibfnamefont {J.}~\bibnamefont {Mangeney}},\ }\bibfield  {title}
  {\bibinfo {title} {Thz spectroscopy for fundamental science and
  applications},\ }\href {https://doi.org/10.1051/photon/202010133} {\bibfield
  {journal} {\bibinfo  {journal} {Photoniques}\ }\textbf {\bibinfo {volume}
  {101}},\ \bibinfo {pages} {33} (\bibinfo {year} {2020})}\BibitemShut
  {NoStop}%
  \bibitem [{\citenamefont {Helm}\ \emph {et~al.}(2023)\citenamefont {Helm},
  \citenamefont {Winnerl}, \citenamefont {Pashkin}, \citenamefont {Klopf},
  \citenamefont {Deinert}, \citenamefont {Kovalev}, \citenamefont {Evtushenko},
  \citenamefont {Lehnert}, \citenamefont {Xiang}, \citenamefont {Arnold} \emph
  {et~al.}}]{helm2023elbe}%
  \BibitemOpen
  \bibfield  {author} {\bibinfo {author} {\bibfnamefont {M.}~\bibnamefont
  {Helm}}, \bibinfo {author} {\bibfnamefont {S.}~\bibnamefont {Winnerl}},
  \bibinfo {author} {\bibfnamefont {A.}~\bibnamefont {Pashkin}}, \bibinfo
  {author} {\bibfnamefont {J.~M.}\ \bibnamefont {Klopf}}, \bibinfo {author}
  {\bibfnamefont {J.-C.}\ \bibnamefont {Deinert}}, \bibinfo {author}
  {\bibfnamefont {S.}~\bibnamefont {Kovalev}}, \bibinfo {author} {\bibfnamefont
  {P.}~\bibnamefont {Evtushenko}}, \bibinfo {author} {\bibfnamefont
  {U.}~\bibnamefont {Lehnert}}, \bibinfo {author} {\bibfnamefont
  {R.}~\bibnamefont {Xiang}}, \bibinfo {author} {\bibfnamefont
  {A.}~\bibnamefont {Arnold}}, \emph {et~al.},\ }\bibfield  {title} {\bibinfo
  {title} {{The Elbe infrared and THz facility at Helmholtz-zentrum
  Dresden-Rossendorf}},\ }\href@noop {} {\bibfield  {journal} {\bibinfo
  {journal} {The European Physical Journal Plus}\ }\textbf {\bibinfo {volume}
  {138}},\ \bibinfo {pages} {158} (\bibinfo {year} {2023})}\BibitemShut
  {NoStop}%
\bibitem [{\citenamefont {Lu}\ \emph {et~al.}(2024)\citenamefont {Lu},
  \citenamefont {Huang}, \citenamefont {Cheng}, \citenamefont {Ma},
  \citenamefont {Xu}, \citenamefont {Zang}, \citenamefont {Wu},\ and\
  \citenamefont {Xu}}]{Lu2024}%
  \BibitemOpen
  \bibfield  {author} {\bibinfo {author} {\bibfnamefont {Y.}~\bibnamefont
  {Lu}}, \bibinfo {author} {\bibfnamefont {Y.}~\bibnamefont {Huang}}, \bibinfo
  {author} {\bibfnamefont {J.}~\bibnamefont {Cheng}}, \bibinfo {author}
  {\bibfnamefont {R.}~\bibnamefont {Ma}}, \bibinfo {author} {\bibfnamefont
  {X.}~\bibnamefont {Xu}}, \bibinfo {author} {\bibfnamefont {Y.}~\bibnamefont
  {Zang}}, \bibinfo {author} {\bibfnamefont {Q.}~\bibnamefont {Wu}},\ and\
  \bibinfo {author} {\bibfnamefont {J.}~\bibnamefont {Xu}},\ }\bibfield
  {title} {\bibinfo {title} {Nonlinear optical physics at terahertz
  frequency},\ }\href {https://doi.org/10.1515/nanoph-2024-0109} {\bibfield
  {journal} {\bibinfo  {journal} {Nanophotonics}\ }\textbf {\bibinfo {volume}
  {13}},\ \bibinfo {pages} {3279--3298} (\bibinfo {year} {2024})}\BibitemShut
  {NoStop}%
\bibitem [{\citenamefont {Lein{\ss}}\ \emph {et~al.}(2008)\citenamefont
  {Lein{\ss}}, \citenamefont {Kampfrath}, \citenamefont {v.~Volkmann},
  \citenamefont {Wolf}, \citenamefont {Steiner}, \citenamefont {Kira},
  \citenamefont {Koch}, \citenamefont {Leitenstorfer},\ and\ \citenamefont
  {Huber}}]{leinss2008terahertz}%
  \BibitemOpen
  \bibfield  {author} {\bibinfo {author} {\bibfnamefont {S.}~\bibnamefont
  {Lein{\ss}}}, \bibinfo {author} {\bibfnamefont {T.}~\bibnamefont
  {Kampfrath}}, \bibinfo {author} {\bibfnamefont {K.}~\bibnamefont
  {v.~Volkmann}}, \bibinfo {author} {\bibfnamefont {M.}~\bibnamefont {Wolf}},
  \bibinfo {author} {\bibfnamefont {J.~T.}\ \bibnamefont {Steiner}}, \bibinfo
  {author} {\bibfnamefont {M.}~\bibnamefont {Kira}}, \bibinfo {author}
  {\bibfnamefont {S.~W.}\ \bibnamefont {Koch}}, \bibinfo {author}
  {\bibfnamefont {A.}~\bibnamefont {Leitenstorfer}},\ and\ \bibinfo {author}
  {\bibfnamefont {R.}~\bibnamefont {Huber}},\ }\bibfield  {title} {\bibinfo
  {title} {Terahertz coherent control of optically dark paraexcitons in
  {Cu$_2$O}},\ }\href@noop {} {\bibfield  {journal} {\bibinfo  {journal}
  {Physical Review Letters}\ }\textbf {\bibinfo {volume} {101}},\ \bibinfo
  {pages} {246401} (\bibinfo {year} {2008})}\BibitemShut {NoStop}%
\bibitem [{\citenamefont {Venanzi}\ \emph {et~al.}(2021)\citenamefont
  {Venanzi}, \citenamefont {Selig}, \citenamefont {Winnerl}, \citenamefont
  {Pashkin}, \citenamefont {Knorr}, \citenamefont {Helm},\ and\ \citenamefont
  {Schneider}}]{venanzi2021terahertz}%
  \BibitemOpen
  \bibfield  {author} {\bibinfo {author} {\bibfnamefont {T.}~\bibnamefont
  {Venanzi}}, \bibinfo {author} {\bibfnamefont {M.}~\bibnamefont {Selig}},
  \bibinfo {author} {\bibfnamefont {S.}~\bibnamefont {Winnerl}}, \bibinfo
  {author} {\bibfnamefont {A.}~\bibnamefont {Pashkin}}, \bibinfo {author}
  {\bibfnamefont {A.}~\bibnamefont {Knorr}}, \bibinfo {author} {\bibfnamefont
  {M.}~\bibnamefont {Helm}},\ and\ \bibinfo {author} {\bibfnamefont
  {H.}~\bibnamefont {Schneider}},\ }\bibfield  {title} {\bibinfo {title}
  {Terahertz-induced energy transfer from hot carriers to trions in a
  {MoSe$_2$} monolayer},\ }\href@noop {} {\bibfield  {journal} {\bibinfo
  {journal} {ACS Photonics}\ }\textbf {\bibinfo {volume} {8}},\ \bibinfo
  {pages} {2931} (\bibinfo {year} {2021})}\BibitemShut {NoStop}%
  \bibitem{Ashkinadze:1988aa}
B.~M. Ashkinadze, V.~V. Belkov, and A.~V. Subashiev, Optical and electrical bistability induced by exciton ionization
  processes, physica status solidi (b) {\bf 150}, 533 (1988).
\bibitem [{\citenamefont {Chen-Esterlit}\ \emph {et~al.}(1996)\citenamefont
  {Chen-Esterlit}, \citenamefont {Lifshitz}, \citenamefont {Cohen},\ and\
  \citenamefont {Pfeiffer}}]{PhysRevB.53.10921}%
  \BibitemOpen
  \bibfield  {author} {\bibinfo {author} {\bibfnamefont {Z.}~\bibnamefont
  {Chen-Esterlit}}, \bibinfo {author} {\bibfnamefont {E.}~\bibnamefont
  {Lifshitz}}, \bibinfo {author} {\bibfnamefont {E.}~\bibnamefont {Cohen}},\
  and\ \bibinfo {author} {\bibfnamefont {L.~N.}\ \bibnamefont {Pfeiffer}},\
  }\bibfield  {title} {\bibinfo {title} {Microwave modulation of circularly
  polarized exciton photonluminescence in {GaAs/AlAs} multiple quantum wells},\
  }\href {https://doi.org/10.1103/PhysRevB.53.10921} {\bibfield  {journal}
  {\bibinfo  {journal} {Phys. Rev. B}\ }\textbf {\bibinfo {volume} {53}},\
  \bibinfo {pages} {10921} (\bibinfo {year} {1996})}\BibitemShut {NoStop}%
\bibitem [{\citenamefont {Lifshitz}\ and\ \citenamefont
  {Bykov}(1993)}]{Lifshitz1993}%
  \BibitemOpen
  \bibfield  {author} {\bibinfo {author} {\bibfnamefont {E.}~\bibnamefont
  {Lifshitz}}\ and\ \bibinfo {author} {\bibfnamefont {L.}~\bibnamefont
  {Bykov}},\ }\bibfield  {title} {\bibinfo {title} {Microwave modulated and
  thermal modulated photoluminescence studies of {2H}-lead iodide},\ }\href
  {https://doi.org/10.1021/j100139a006} {\bibfield  {journal} {\bibinfo
  {journal} {The Journal of Physical Chemistry}\ }\textbf {\bibinfo {volume}
  {97}},\ \bibinfo {pages} {9288} (\bibinfo {year} {1993})}\BibitemShut
  {NoStop}%
\bibitem [{\citenamefont {Venanzi}\ \emph {et~al.}(2024)\citenamefont
  {Venanzi}, \citenamefont {Cuccu}, \citenamefont {Perea-Causin}, \citenamefont
  {Sun}, \citenamefont {Brem}, \citenamefont {Erkensten}, \citenamefont
  {Taniguchi}, \citenamefont {Watanabe}, \citenamefont {Malic}, \citenamefont
  {Helm} \emph {et~al.}}]{venanzi2024ultrafast}%
  \BibitemOpen
  \bibfield  {author} {\bibinfo {author} {\bibfnamefont {T.}~\bibnamefont
  {Venanzi}}, \bibinfo {author} {\bibfnamefont {M.}~\bibnamefont {Cuccu}},
  \bibinfo {author} {\bibfnamefont {R.}~\bibnamefont {Perea-Causin}}, \bibinfo
  {author} {\bibfnamefont {X.}~\bibnamefont {Sun}}, \bibinfo {author}
  {\bibfnamefont {S.}~\bibnamefont {Brem}}, \bibinfo {author} {\bibfnamefont
  {D.}~\bibnamefont {Erkensten}}, \bibinfo {author} {\bibfnamefont
  {T.}~\bibnamefont {Taniguchi}}, \bibinfo {author} {\bibfnamefont
  {K.}~\bibnamefont {Watanabe}}, \bibinfo {author} {\bibfnamefont
  {E.}~\bibnamefont {Malic}}, \bibinfo {author} {\bibfnamefont
  {M.}~\bibnamefont {Helm}}, {\bibfnamefont
  {S.}~\bibnamefont {Winnerl}}, {\bibfnamefont
  {A.}~\bibnamefont {Chernikov}},\ }\bibfield  {title} {\bibinfo
  {title} {Ultrafast switching of trions in {2D} materials by terahertz
  photons},\ }\href@noop {} {\bibfield  {journal} {\bibinfo  {journal} {Nature
  Photonics}\ }\textbf {\bibinfo {volume} {18}},\ \bibinfo {pages} {1344}
  (\bibinfo {year} {2024})}\BibitemShut {NoStop}%
\bibitem [{\citenamefont {Ossau}\ and\ \citenamefont
  {Suris}(2003)}]{suris:correlation}%
  \BibitemOpen
  \bibinfo {editor} {\bibfnamefont {W.}~\bibnamefont {Ossau}}\ and\ \bibinfo
  {editor} {\bibfnamefont {R.}~\bibnamefont {Suris}},\ eds.,\ \bibinfo {title}
  {Optical properties of 2d systems with interacting electrons}\ (\bibinfo
  {publisher} {NATO ASI},\ \bibinfo {year} {2003})\ Chap.\ \bibinfo {chapter}
  {R. A. Suris, {Correlation between trion and hole in Fermi distribution in
  process of trion photo-excitation in doped QWs}}\BibitemShut {NoStop}%
\bibitem [{\citenamefont {Koudinov}\ \emph {et~al.}(2014)\citenamefont
  {Koudinov}, \citenamefont {Kehl}, \citenamefont {Rodina}, \citenamefont
  {Geurts}, \citenamefont {Wolverson},\ and\ \citenamefont
  {Karczewski}}]{PhysRevLett.112.147402}%
  \BibitemOpen
  \bibfield  {author} {\bibinfo {author} {\bibfnamefont {A.~V.}\ \bibnamefont
  {Koudinov}}, \bibinfo {author} {\bibfnamefont {C.}~\bibnamefont {Kehl}},
  \bibinfo {author} {\bibfnamefont {A.~V.}\ \bibnamefont {Rodina}}, \bibinfo
  {author} {\bibfnamefont {J.}~\bibnamefont {Geurts}}, \bibinfo {author}
  {\bibfnamefont {D.}~\bibnamefont {Wolverson}},\ and\ \bibinfo {author}
  {\bibfnamefont {G.}~\bibnamefont {Karczewski}},\ }\bibfield  {title}
  {\bibinfo {title} {Suris tetrons: Possible spectroscopic evidence for
  four-particle optical excitations of a two-dimensional electron gas},\ }\href
  {https://doi.org/10.1103/PhysRevLett.112.147402} {\bibfield  {journal}
  {\bibinfo  {journal} {Phys. Rev. Lett.}\ }\textbf {\bibinfo {volume} {112}},\
  \bibinfo {pages} {147402} (\bibinfo {year} {2014})}\BibitemShut {NoStop}%
\bibitem [{\citenamefont {Rapaport}\ \emph {et~al.}(2001)\citenamefont
  {Rapaport}, \citenamefont {Cohen}, \citenamefont {Ron}, \citenamefont
  {Linder},\ and\ \citenamefont {Pfeiffer}}]{PhysRevB.63.235310}%
  \BibitemOpen
  \bibfield  {author} {\bibinfo {author} {\bibfnamefont {R.}~\bibnamefont
  {Rapaport}}, \bibinfo {author} {\bibfnamefont {E.}~\bibnamefont {Cohen}},
  \bibinfo {author} {\bibfnamefont {A.}~\bibnamefont {Ron}}, \bibinfo {author}
  {\bibfnamefont {E.}~\bibnamefont {Linder}},\ and\ \bibinfo {author}
  {\bibfnamefont {L.~N.}\ \bibnamefont {Pfeiffer}},\ }\bibfield  {title}
  {\bibinfo {title} {Negatively charged polaritons in a semiconductor
  microcavity},\ }\href {https://doi.org/10.1103/PhysRevB.63.235310} {\bibfield
   {journal} {\bibinfo  {journal} {Phys. Rev. B}\ }\textbf {\bibinfo {volume}
  {63}},\ \bibinfo {pages} {235310} (\bibinfo {year} {2001})}\BibitemShut
  {NoStop}%
\bibitem [{\citenamefont {Sidler}\ \emph {et~al.}(2016)\citenamefont {Sidler},
  \citenamefont {Back}, \citenamefont {Cotlet}, \citenamefont {Srivastava},
  \citenamefont {Fink}, \citenamefont {Kroner}, \citenamefont {Demler},\ and\
  \citenamefont {Imamoglu}}]{Sidler:2016aa}%
  \BibitemOpen
  \bibfield  {author} {\bibinfo {author} {\bibfnamefont {M.}~\bibnamefont
  {Sidler}}, \bibinfo {author} {\bibfnamefont {P.}~\bibnamefont {Back}},
  \bibinfo {author} {\bibfnamefont {O.}~\bibnamefont {Cotlet}}, \bibinfo
  {author} {\bibfnamefont {A.}~\bibnamefont {Srivastava}}, \bibinfo {author}
  {\bibfnamefont {T.}~\bibnamefont {Fink}}, \bibinfo {author} {\bibfnamefont
  {M.}~\bibnamefont {Kroner}}, \bibinfo {author} {\bibfnamefont
  {E.}~\bibnamefont {Demler}},\ and\ \bibinfo {author} {\bibfnamefont
  {A.}~\bibnamefont {Imamoglu}},\ }\bibfield  {title} {\bibinfo {title} {Fermi
  polaron-polaritons in charge-tunable atomically thin semiconductors},\ }\href
  {http://dx.doi.org/10.1038/nphys3949} {\bibfield  {journal} {\bibinfo
  {journal} {Nature Physics}\ }\textbf {\bibinfo {volume} {13}},\ \bibinfo
  {pages} {255} (\bibinfo {year} {2016})}\BibitemShut {NoStop}%
\bibitem [{\citenamefont {Efimkin}\ and\ \citenamefont
  {MacDonald}(2017)}]{PhysRevB.95.035417}%
  \BibitemOpen
  \bibfield  {author} {\bibinfo {author} {\bibfnamefont {D.~K.}\ \bibnamefont
  {Efimkin}}\ and\ \bibinfo {author} {\bibfnamefont {A.~H.}\ \bibnamefont
  {MacDonald}},\ }\bibfield  {title} {\bibinfo {title} {Many-body theory of
  trion absorption features in two-dimensional semiconductors},\ }\href
  {https://doi.org/10.1103/PhysRevB.95.035417} {\bibfield  {journal} {\bibinfo
  {journal} {Phys. Rev. B}\ }\textbf {\bibinfo {volume} {95}},\ \bibinfo
  {pages} {035417} (\bibinfo {year} {2017})}\BibitemShut {NoStop}%
\bibitem [{\citenamefont {Rana}\ \emph {et~al.}(2020)\citenamefont {Rana},
  \citenamefont {Koksal},\ and\ \citenamefont
  {Manolatou}}]{PhysRevB.102.085304}%
  \BibitemOpen
  \bibfield  {author} {\bibinfo {author} {\bibfnamefont {F.}~\bibnamefont
  {Rana}}, \bibinfo {author} {\bibfnamefont {O.}~\bibnamefont {Koksal}},\ and\
  \bibinfo {author} {\bibfnamefont {C.}~\bibnamefont {Manolatou}},\ }\bibfield
  {title} {\bibinfo {title} {Many-body theory of the optical conductivity of
  excitons and trions in two-dimensional materials},\ }\href
  {https://doi.org/10.1103/PhysRevB.102.085304} {\bibfield  {journal} {\bibinfo
   {journal} {Phys. Rev. B}\ }\textbf {\bibinfo {volume} {102}},\ \bibinfo
  {pages} {085304} (\bibinfo {year} {2020})}\BibitemShut {NoStop}%
\bibitem [{\citenamefont {Mulkerin}\ \emph {et~al.}(2023)\citenamefont
  {Mulkerin}, \citenamefont {Tiene}, \citenamefont {Marchetti}, \citenamefont
  {Parish},\ and\ \citenamefont {Levinsen}}]{PhysRevLett.131.106901}%
  \BibitemOpen
  \bibfield  {author} {\bibinfo {author} {\bibfnamefont {B.~C.}\ \bibnamefont
  {Mulkerin}}, \bibinfo {author} {\bibfnamefont {A.}~\bibnamefont {Tiene}},
  \bibinfo {author} {\bibfnamefont {F.~M.}\ \bibnamefont {Marchetti}}, \bibinfo
  {author} {\bibfnamefont {M.~M.}\ \bibnamefont {Parish}},\ and\ \bibinfo
  {author} {\bibfnamefont {J.}~\bibnamefont {Levinsen}},\ }\bibfield  {title}
  {\bibinfo {title} {Exact quantum virial expansion for the optical response of
  doped two-dimensional semiconductors},\ }\href
  {https://doi.org/10.1103/PhysRevLett.131.106901} {\bibfield  {journal}
  {\bibinfo  {journal} {Phys. Rev. Lett.}\ }\textbf {\bibinfo {volume} {131}},\
  \bibinfo {pages} {106901} (\bibinfo {year} {2023})}\BibitemShut {NoStop}%
\bibitem [{\citenamefont {Tiene}\ \emph {et~al.}(2023)\citenamefont {Tiene},
  \citenamefont {Mulkerin}, \citenamefont {Levinsen}, \citenamefont {Parish},\
  and\ \citenamefont {Marchetti}}]{PhysRevB.108.125406}%
  \BibitemOpen
  \bibfield  {author} {\bibinfo {author} {\bibfnamefont {A.}~\bibnamefont
  {Tiene}}, \bibinfo {author} {\bibfnamefont {B.~C.}\ \bibnamefont {Mulkerin}},
  \bibinfo {author} {\bibfnamefont {J.}~\bibnamefont {Levinsen}}, \bibinfo
  {author} {\bibfnamefont {M.~M.}\ \bibnamefont {Parish}},\ and\ \bibinfo
  {author} {\bibfnamefont {F.~M.}\ \bibnamefont {Marchetti}},\ }\bibfield
  {title} {\bibinfo {title} {Crossover from exciton polarons to trions in doped
  two-dimensional semiconductors at finite temperature},\ }\href
  {https://doi.org/10.1103/PhysRevB.108.125406} {\bibfield  {journal} {\bibinfo
   {journal} {Phys. Rev. B}\ }\textbf {\bibinfo {volume} {108}},\ \bibinfo
  {pages} {125406} (\bibinfo {year} {2023})}\BibitemShut {NoStop}%
\bibitem [{\citenamefont {Glazov}(2020)}]{Glazov:2020wf}%
  \BibitemOpen
  \bibfield  {author} {\bibinfo {author} {\bibfnamefont {M.~M.}\ \bibnamefont
  {Glazov}},\ }\bibfield  {title} {\bibinfo {title} {Optical properties of
  charged excitons in two-dimensional semiconductors},\ }\href
  {https://doi.org/10.1063/5.0012475} {\bibfield  {journal} {\bibinfo
  {journal} {J. Chem. Phys.}\ }\textbf {\bibinfo {volume} {153}},\ \bibinfo
  {pages} {034703} (\bibinfo {year} {2020})}\BibitemShut {NoStop}%
\bibitem [{\citenamefont {Zipfel}\ \emph {et~al.}(2022)\citenamefont {Zipfel},
  \citenamefont {Wagner}, \citenamefont {Semina}, \citenamefont {Ziegler},
  \citenamefont {Taniguchi}, \citenamefont {Watanabe}, \citenamefont {Glazov},\
  and\ \citenamefont {Chernikov}}]{PhysRevB.105.075311}%
  \BibitemOpen
  \bibfield  {author} {\bibinfo {author} {\bibfnamefont {J.}~\bibnamefont
  {Zipfel}}, \bibinfo {author} {\bibfnamefont {K.}~\bibnamefont {Wagner}},
  \bibinfo {author} {\bibfnamefont {M.~A.}\ \bibnamefont {Semina}}, \bibinfo
  {author} {\bibfnamefont {J.~D.}\ \bibnamefont {Ziegler}}, \bibinfo {author}
  {\bibfnamefont {T.}~\bibnamefont {Taniguchi}}, \bibinfo {author}
  {\bibfnamefont {K.}~\bibnamefont {Watanabe}}, \bibinfo {author}
  {\bibfnamefont {M.~M.}\ \bibnamefont {Glazov}},\ and\ \bibinfo {author}
  {\bibfnamefont {A.}~\bibnamefont {Chernikov}},\ }\bibfield  {title} {\bibinfo
  {title} {Electron recoil effect in electrically tunable {MoSe$_{2}$}
  monolayers},\ }\href {https://doi.org/10.1103/PhysRevB.105.075311} {\bibfield
   {journal} {\bibinfo  {journal} {Phys. Rev. B}\ }\textbf {\bibinfo {volume}
  {105}},\ \bibinfo {pages} {075311} (\bibinfo {year} {2022})}\BibitemShut
  {NoStop}%
\bibitem [{\citenamefont {Iakovlev}\ and\ \citenamefont
  {Glazov}(2023)}]{iakovlev2023fermi}%
  \BibitemOpen
  \bibfield  {author} {\bibinfo {author} {\bibfnamefont {Z.~A.}\ \bibnamefont
  {Iakovlev}}\ and\ \bibinfo {author} {\bibfnamefont {M.~M.}\ \bibnamefont
  {Glazov}},\ }\bibfield  {title} {\bibinfo {title} {Fermi polaron fine
  structure in strained van der {Waals} heterostructures},\ }\href@noop {}
  {\bibfield  {journal} {\bibinfo  {journal} {2D Materials}\ }\textbf {\bibinfo
  {volume} {10}},\ \bibinfo {pages} {035034} (\bibinfo {year}
  {2023})}\BibitemShut {NoStop}%
\bibitem [{\citenamefont {Iakovlev}\ and\ \citenamefont
  {Glazov}(2024)}]{iakovlev2024longitudinal}%
  \BibitemOpen
  \bibfield  {author} {\bibinfo {author} {\bibfnamefont {Z.~A.}\ \bibnamefont
  {Iakovlev}}\ and\ \bibinfo {author} {\bibfnamefont {M.~M.}\ \bibnamefont
  {Glazov}},\ }\bibfield  {title} {\bibinfo {title} {Longitudinal-transverse
  splitting and fine structure of {Fermi} polarons in two-dimensional
  semiconductors},\ }\href@noop {} {\bibfield  {journal} {\bibinfo  {journal}
  {Journal of Luminescence}\ }\textbf {\bibinfo {volume} {273}},\ \bibinfo
  {pages} {120700} (\bibinfo {year} {2024})}\BibitemShut {NoStop}%
\bibitem [{\citenamefont {Yagodkin}\ \emph {et~al.}(2024)\citenamefont
  {Yagodkin}, \citenamefont {Burfeindt}, \citenamefont {Iakovlev},
  \citenamefont {Kumar}, \citenamefont {Dewambrechies}, \citenamefont
  {Y{\"u}cel}, \citenamefont {H{\"o}fer}, \citenamefont {Gahl}, \citenamefont
  {Glazov},\ and\ \citenamefont
  {Bolotin}}]{yagodkin2024excitonslargepseudomagneticfields}%
  \BibitemOpen
  \bibfield  {author} {\bibinfo {author} {\bibfnamefont {D.}~\bibnamefont
  {Yagodkin}}, \bibinfo {author} {\bibfnamefont {K.}~\bibnamefont {Burfeindt}},
  \bibinfo {author} {\bibfnamefont {Z.~A.}\ \bibnamefont {Iakovlev}}, \bibinfo
  {author} {\bibfnamefont {A.~M.}\ \bibnamefont {Kumar}}, \bibinfo {author}
  {\bibfnamefont {A.}~\bibnamefont {Dewambrechies}}, \bibinfo {author}
  {\bibfnamefont {O.}~\bibnamefont {Y{\"u}cel}}, \bibinfo {author}
  {\bibfnamefont {B.}~\bibnamefont {H{\"o}fer}}, \bibinfo {author}
  {\bibfnamefont {C.}~\bibnamefont {Gahl}}, \bibinfo {author} {\bibfnamefont
  {M.~M.}\ \bibnamefont {Glazov}},\ and\ \bibinfo {author} {\bibfnamefont
  {K.~I.}\ \bibnamefont {Bolotin}},\ }\href {https://arxiv.org/abs/2412.16596}
  {\bibinfo {title} {Excitons under large pseudomagnetic fields}},  \Eprint {https://arxiv.org/abs/2412.16596} {arXiv:2412.16596} (\bibinfo
  {year} {2024}) \BibitemShut {NoStop}%
\bibitem [{\citenamefont {Schmidt}\ \emph {et~al.}(2018)\citenamefont
  {Schmidt}, \citenamefont {Knap}, \citenamefont {Ivanov}, \citenamefont {You},
  \citenamefont {Cetina},\ and\ \citenamefont {Demler}}]{Schmidt:2018aa}%
  \BibitemOpen
  \bibfield  {author} {\bibinfo {author} {\bibfnamefont {R.}~\bibnamefont
  {Schmidt}}, \bibinfo {author} {\bibfnamefont {M.}~\bibnamefont {Knap}},
  \bibinfo {author} {\bibfnamefont {D.~A.}\ \bibnamefont {Ivanov}}, \bibinfo
  {author} {\bibfnamefont {J.-S.}\ \bibnamefont {You}}, \bibinfo {author}
  {\bibfnamefont {M.}~\bibnamefont {Cetina}},\ and\ \bibinfo {author}
  {\bibfnamefont {E.}~\bibnamefont {Demler}},\ }\bibfield  {title} {\bibinfo
  {title} {Universal many-body response of heavy impurities coupled to a
  {Fermi} sea: a review of recent progress},\ }\href
  {https://doi.org/10.1088/1361-6633/aa9593} {\bibfield  {journal} {\bibinfo
  {journal} {Reports on Progress in Physics}\ }\textbf {\bibinfo {volume}
  {81}},\ \bibinfo {pages} {024401} (\bibinfo {year} {2018})}\BibitemShut
  {NoStop}%
\bibitem [{\citenamefont {Fey}\ \emph {et~al.}(2020)\citenamefont {Fey},
  \citenamefont {Schmelcher}, \citenamefont {Imamoglu},\ and\ \citenamefont
  {Schmidt}}]{2019arXiv191204873F}%
  \BibitemOpen
  \bibfield  {author} {\bibinfo {author} {\bibfnamefont {C.}~\bibnamefont
  {Fey}}, \bibinfo {author} {\bibfnamefont {P.}~\bibnamefont {Schmelcher}},
  \bibinfo {author} {\bibfnamefont {A.}~\bibnamefont {Imamoglu}},\ and\
  \bibinfo {author} {\bibfnamefont {R.}~\bibnamefont {Schmidt}},\ }\bibfield
  {title} {\bibinfo {title} {Theory of exciton-electron scattering in
  atomically thin semiconductors},\ }\href
  {https://doi.org/10.1103/PhysRevB.101.195417} {\bibfield  {journal} {\bibinfo
   {journal} {Phys. Rev. B}\ }\textbf {\bibinfo {volume} {101}},\ \bibinfo
  {pages} {195417} (\bibinfo {year} {2020})}\BibitemShut {NoStop}%
\bibitem [{\citenamefont {Chevy}(2006)}]{PhysRevA.74.063628}%
  \BibitemOpen
  \bibfield  {author} {\bibinfo {author} {\bibfnamefont {F.}~\bibnamefont
  {Chevy}},\ }\bibfield  {title} {\bibinfo {title} {Universal phase diagram of
  a strongly interacting fermi gas with unbalanced spin populations},\ }\href
  {https://doi.org/10.1103/PhysRevA.74.063628} {\bibfield  {journal} {\bibinfo
  {journal} {Phys. Rev. A}\ }\textbf {\bibinfo {volume} {74}},\ \bibinfo
  {pages} {063628} (\bibinfo {year} {2006})}\BibitemShut {NoStop}%
\bibitem [{\citenamefont {Naftaly}\ \emph {et~al.}(2011)\citenamefont
  {Naftaly}, \citenamefont {Leist},\ and\ \citenamefont
  {Dudley}}]{naftaly2011hexagonal}%
  \BibitemOpen
  \bibfield  {author} {\bibinfo {author} {\bibfnamefont {M.}~\bibnamefont
  {Naftaly}}, \bibinfo {author} {\bibfnamefont {J.}~\bibnamefont {Leist}},\
  and\ \bibinfo {author} {\bibfnamefont {R.}~\bibnamefont {Dudley}},\
  }\bibfield  {title} {\bibinfo {title} {Hexagonal boron nitride studied by
  terahertz time-domain spectroscopy},\ }\href@noop {} {\bibfield  {journal}
  {\bibinfo  {journal} {Journal of Physics: Conference Series}\ }\textbf
  {\bibinfo {volume} {310}},\ \bibinfo {pages} {012006} (\bibinfo {year}
  {2011})}\BibitemShut {NoStop}%
\bibitem [{\citenamefont {Wagner}\ \emph {et~al.}(2023)\citenamefont {Wagner},
  \citenamefont {Iakovlev}, \citenamefont {Ziegler}, \citenamefont {Cuccu},
  \citenamefont {Taniguchi}, \citenamefont {Watanabe}, \citenamefont {Glazov},\
  and\ \citenamefont {Chernikov}}]{wagner:trions}%
  \BibitemOpen
  \bibfield  {author} {\bibinfo {author} {\bibfnamefont {K.}~\bibnamefont
  {Wagner}}, \bibinfo {author} {\bibfnamefont {Z.~A.}\ \bibnamefont
  {Iakovlev}}, \bibinfo {author} {\bibfnamefont {J.~D.}\ \bibnamefont
  {Ziegler}}, \bibinfo {author} {\bibfnamefont {M.}~\bibnamefont {Cuccu}},
  \bibinfo {author} {\bibfnamefont {T.}~\bibnamefont {Taniguchi}}, \bibinfo
  {author} {\bibfnamefont {K.}~\bibnamefont {Watanabe}}, \bibinfo {author}
  {\bibfnamefont {M.~M.}\ \bibnamefont {Glazov}},\ and\ \bibinfo {author}
  {\bibfnamefont {A.}~\bibnamefont {Chernikov}},\ }\bibfield  {title} {\bibinfo
  {title} {Diffusion of excitons in a two-dimensional {F}ermi sea of free
  charges},\ }\href {https://doi.org/10.1021/acs.nanolett.2c03796} {\bibfield
  {journal} {\bibinfo  {journal} {Nano Letters}\ }\textbf {\bibinfo {volume}
  {23}},\ \bibinfo {pages} {4708} (\bibinfo {year} {2023})}\BibitemShut
  {NoStop}%
\bibitem [{\citenamefont {Budkin}\ and\ \citenamefont
  {Tarasenko}(2011)}]{Budkin:2011aa}%
  \BibitemOpen
  \bibfield  {author} {\bibinfo {author} {\bibfnamefont {G.~V.}\ \bibnamefont
  {Budkin}}\ and\ \bibinfo {author} {\bibfnamefont {S.~A.}\ \bibnamefont
  {Tarasenko}},\ }\bibfield  {title} {\bibinfo {title} {Heating and cooling of
  a two-dimensional electron gas by terahertz radiation},\ }\href
  {https://doi.org/10.1134/S1063776111030046} {\bibfield  {journal} {\bibinfo
  {journal} {JETP}\ }\textbf {\bibinfo {volume} {112}},\ \bibinfo {pages} {656}
  (\bibinfo {year} {2011})}\BibitemShut {NoStop}%
\bibitem [{\citenamefont {Kaasbjerg}\ \emph {et~al.}(2012)\citenamefont
  {Kaasbjerg}, \citenamefont {Thygesen},\ and\ \citenamefont
  {Jacobsen}}]{kaasbjerg2012phonon}%
  \BibitemOpen
  \bibfield  {author} {\bibinfo {author} {\bibfnamefont {K.}~\bibnamefont
  {Kaasbjerg}}, \bibinfo {author} {\bibfnamefont {K.~S.}\ \bibnamefont
  {Thygesen}},\ and\ \bibinfo {author} {\bibfnamefont {K.~W.}\ \bibnamefont
  {Jacobsen}},\ }\bibfield  {title} {\bibinfo {title} {Phonon-limited mobility
  in n-type single-layer {MoS$_2$} from first principles},\ }\href@noop {}
  {\bibfield  {journal} {\bibinfo  {journal} {Physical Review B}\ }\textbf
  {\bibinfo {volume} {85}},\ \bibinfo {pages} {115317} (\bibinfo {year}
  {2012})}\BibitemShut {NoStop}%
\bibitem [{\citenamefont {Kaasbjerg}\ \emph {et~al.}(2013)\citenamefont
  {Kaasbjerg}, \citenamefont {Thygesen},\ and\ \citenamefont
  {Jauho}}]{kaasbjerg2013acoustic}%
  \BibitemOpen
  \bibfield  {author} {\bibinfo {author} {\bibfnamefont {K.}~\bibnamefont
  {Kaasbjerg}}, \bibinfo {author} {\bibfnamefont {K.~S.}\ \bibnamefont
  {Thygesen}},\ and\ \bibinfo {author} {\bibfnamefont {A.-P.}\ \bibnamefont
  {Jauho}},\ }\bibfield  {title} {\bibinfo {title} {Acoustic phonon limited
  mobility in two-dimensional semiconductors: Deformation potential and
  piezoelectric scattering in monolayer {MoS$_2$} from first principles},\
  }\href@noop {} {\bibfield  {journal} {\bibinfo  {journal} {Physical Review
  B}\ }\textbf {\bibinfo {volume} {87}},\ \bibinfo {pages} {235312} (\bibinfo
  {year} {2013})}\BibitemShut {NoStop}%
\bibitem [{\citenamefont {Kaasbjerg}\ \emph {et~al.}(2014)\citenamefont
  {Kaasbjerg}, \citenamefont {Bhargavi},\ and\ \citenamefont
  {Kubakaddi}}]{kaasbjerg2014hot}%
  \BibitemOpen
  \bibfield  {author} {\bibinfo {author} {\bibfnamefont {K.}~\bibnamefont
  {Kaasbjerg}}, \bibinfo {author} {\bibfnamefont {K.}~\bibnamefont
  {Bhargavi}},\ and\ \bibinfo {author} {\bibfnamefont {S.}~\bibnamefont
  {Kubakaddi}},\ }\bibfield  {title} {\bibinfo {title} {Hot-electron cooling by
  acoustic and optical phonons in monolayers of {MoS$_2$} and other
  transition-metal dichalcogenides},\ }\href@noop {} {\bibfield  {journal}
  {\bibinfo  {journal} {Physical Review B}\ }\textbf {\bibinfo {volume} {90}},\
  \bibinfo {pages} {165436} (\bibinfo {year} {2014})}\BibitemShut {NoStop}%
\bibitem [{\citenamefont {Jin}\ \emph {et~al.}(2014)\citenamefont {Jin},
  \citenamefont {Li}, \citenamefont {Mullen},\ and\ \citenamefont
  {Kim}}]{PhysRevB.90.045422}%
  \BibitemOpen
  \bibfield  {author} {\bibinfo {author} {\bibfnamefont {Z.}~\bibnamefont
  {Jin}}, \bibinfo {author} {\bibfnamefont {X.}~\bibnamefont {Li}}, \bibinfo
  {author} {\bibfnamefont {J.~T.}\ \bibnamefont {Mullen}},\ and\ \bibinfo
  {author} {\bibfnamefont {K.~W.}\ \bibnamefont {Kim}},\ }\bibfield  {title}
  {\bibinfo {title} {Intrinsic transport properties of electrons and holes in
  monolayer transition-metal dichalcogenides},\ }\href
  {https://doi.org/10.1103/PhysRevB.90.045422} {\bibfield  {journal} {\bibinfo
  {journal} {Phys. Rev. B}\ }\textbf {\bibinfo {volume} {90}},\ \bibinfo
  {pages} {045422} (\bibinfo {year} {2014})}\BibitemShut {NoStop}%
\bibitem [{\citenamefont {Shree}\ \emph {et~al.}(2018)\citenamefont {Shree},
  \citenamefont {Semina}, \citenamefont {Robert}, \citenamefont {Han},
  \citenamefont {Amand}, \citenamefont {Balocchi}, \citenamefont {Manca},
  \citenamefont {Courtade}, \citenamefont {Marie}, \citenamefont {Taniguchi},
  \citenamefont {Watanabe}, \citenamefont {Glazov},\ and\ \citenamefont
  {Urbaszek}}]{shree2018exciton}%
  \BibitemOpen
  \bibfield  {author} {\bibinfo {author} {\bibfnamefont {S.}~\bibnamefont
  {Shree}}, \bibinfo {author} {\bibfnamefont {M.}~\bibnamefont {Semina}},
  \bibinfo {author} {\bibfnamefont {C.}~\bibnamefont {Robert}}, \bibinfo
  {author} {\bibfnamefont {B.}~\bibnamefont {Han}}, \bibinfo {author}
  {\bibfnamefont {T.}~\bibnamefont {Amand}}, \bibinfo {author} {\bibfnamefont
  {A.}~\bibnamefont {Balocchi}}, \bibinfo {author} {\bibfnamefont
  {M.}~\bibnamefont {Manca}}, \bibinfo {author} {\bibfnamefont
  {E.}~\bibnamefont {Courtade}}, \bibinfo {author} {\bibfnamefont
  {X.}~\bibnamefont {Marie}}, \bibinfo {author} {\bibfnamefont
  {T.}~\bibnamefont {Taniguchi}}, \bibinfo {author} {\bibfnamefont
  {K.}~\bibnamefont {Watanabe}}, \bibinfo {author} {\bibfnamefont {M.~M.}\
  \bibnamefont {Glazov}},\ and\ \bibinfo {author} {\bibfnamefont
  {B.}~\bibnamefont {Urbaszek}},\ }\bibfield  {title} {\bibinfo {title}
  {Observation of exciton-phonon coupling in {MoSe}$_{2}$ monolayers},\ }\href
  {https://doi.org/10.1103/PhysRevB.98.035302} {\bibfield  {journal} {\bibinfo
  {journal} {Phys. Rev. B}\ }\textbf {\bibinfo {volume} {98}},\ \bibinfo
  {pages} {035302} (\bibinfo {year} {2018})}\BibitemShut {NoStop}%
\bibitem{gantmakher87}
V.~F. Gantmakher and Y.~B. Levinson.
\newblock {\em Carrier Scattering in Metals and Semiconductors}.
\newblock North-Holland Publishing Company, 1987.
\bibitem [{\citenamefont {Golub}\ \emph {et~al.}(1998)\citenamefont {Golub},
  \citenamefont {Ivchenko},\ and\ \citenamefont {Tarasenko}}]{tarasenko98}%
  \BibitemOpen
  \bibfield  {author} {\bibinfo {author} {\bibfnamefont {L.~E.}\ \bibnamefont
  {Golub}}, \bibinfo {author} {\bibfnamefont {E.~L.}\ \bibnamefont
  {Ivchenko}},\ and\ \bibinfo {author} {\bibfnamefont {S.~A.}\ \bibnamefont
  {Tarasenko}},\ }\bibfield  {title} {\bibinfo {title} {Interaction of free
  carriers with localized excitons in quantum wells},\ }\href@noop {}
  {\bibfield  {journal} {\bibinfo  {journal} {Solid State Communications}\
  }\textbf {\bibinfo {volume} {108}},\ \bibinfo {pages} {799} (\bibinfo {year}
  {1998})}\BibitemShut {NoStop}%
\bibitem{PhysRevLett.124.166802}
M.~M. Glazov, Quantum interference effect on exciton transport in monolayer
  semiconductors Phys. Rev. Lett. {\bf 124}, 166802 (2020).

\bibitem{PhysRevB.102.125410}
S.~Ayari, S.~Jaziri, R.~Ferreira, and G.~Bastard, Phonon-assisted exciton/trion conversion efficiency in transition
  metal dichalcogenides, Phys. Rev. B {\bf 102}, 125410 (2020).
\bibitem [{\citenamefont {Tabataba-Vakili}\ \emph {et~al.}(2024)\citenamefont
  {Tabataba-Vakili}, \citenamefont {Nguyen}, \citenamefont {Rupp},
  \citenamefont {Mosina}, \citenamefont {Papavasileiou}, \citenamefont
  {Watanabe}, \citenamefont {Taniguchi}, \citenamefont {Maletinsky},
  \citenamefont {Glazov}, \citenamefont {Sofer}, \citenamefont {Baimuratov},\
  and\ \citenamefont {H{\"o}gele}}]{Tabataba-Vakili:2024aa}%
  \BibitemOpen
  \bibfield  {author} {\bibinfo {author} {\bibfnamefont {F.}~\bibnamefont
  {Tabataba-Vakili}}, \bibinfo {author} {\bibfnamefont {H.~P.~G.}\ \bibnamefont
  {Nguyen}}, \bibinfo {author} {\bibfnamefont {A.}~\bibnamefont {Rupp}},
  \bibinfo {author} {\bibfnamefont {K.}~\bibnamefont {Mosina}}, \bibinfo
  {author} {\bibfnamefont {A.}~\bibnamefont {Papavasileiou}}, \bibinfo {author}
  {\bibfnamefont {K.}~\bibnamefont {Watanabe}}, \bibinfo {author}
  {\bibfnamefont {T.}~\bibnamefont {Taniguchi}}, \bibinfo {author}
  {\bibfnamefont {P.}~\bibnamefont {Maletinsky}}, \bibinfo {author}
  {\bibfnamefont {M.~M.}\ \bibnamefont {Glazov}}, \bibinfo {author}
  {\bibfnamefont {Z.}~\bibnamefont {Sofer}}, \bibinfo {author} {\bibfnamefont
  {A.~S.}\ \bibnamefont {Baimuratov}},\ and\ \bibinfo {author} {\bibfnamefont
  {A.}~\bibnamefont {H{\"o}gele}},\ }\bibfield  {title} {\bibinfo {title}
  {Doping-control of excitons and magnetism in few-layer {CrSBr}},\ }\href
  {https://doi.org/10.1038/s41467-024-49048-9} {\bibfield  {journal} {\bibinfo
  {journal} {Nature Communications}\ }\textbf {\bibinfo {volume} {15}},\
  \bibinfo {pages} {4735} (\bibinfo {year} {2024})}\BibitemShut {NoStop}%
\bibitem [{\citenamefont {Semina}\ \emph {et~al.}(2025)\citenamefont {Semina},
  \citenamefont {Tabataba-Vakili}, \citenamefont {Rupp}, \citenamefont
  {Baimuratov}, \citenamefont {H\"ogele},\ and\ \citenamefont
  {Glazov}}]{PhysRevB.111.205301}%
  \BibitemOpen
  \bibfield  {author} {\bibinfo {author} {\bibfnamefont {M.~A.}\ \bibnamefont
  {Semina}}, \bibinfo {author} {\bibfnamefont {F.}~\bibnamefont
  {Tabataba-Vakili}}, \bibinfo {author} {\bibfnamefont {A.}~\bibnamefont
  {Rupp}}, \bibinfo {author} {\bibfnamefont {A.~S.}\ \bibnamefont
  {Baimuratov}}, \bibinfo {author} {\bibfnamefont {A.}~\bibnamefont
  {H\"ogele}},\ and\ \bibinfo {author} {\bibfnamefont {M.~M.}\ \bibnamefont
  {Glazov}},\ }\bibfield  {title} {\bibinfo {title} {Excitons and trions in
  {CrSBr} bilayers},\ }\href {https://doi.org/10.1103/PhysRevB.111.205301}
  {\bibfield  {journal} {\bibinfo  {journal} {Phys. Rev. B}\ }\textbf {\bibinfo
  {volume} {111}},\ \bibinfo {pages} {205301} (\bibinfo {year}
  {2025})}\BibitemShut {NoStop}%
\bibitem [{\citenamefont {Cotle{\c{t}}}\ \emph {et~al.}(2019)\citenamefont
  {Cotle{\c{t}}}, \citenamefont {Pientka}, \citenamefont {Schmidt},
  \citenamefont {Zarand}, \citenamefont {Demler},\ and\ \citenamefont
  {Imamoglu}}]{cotlect2019transport}%
  \BibitemOpen
  \bibfield  {author} {\bibinfo {author} {\bibfnamefont {O.}~\bibnamefont
  {Cotle{\c{t}}}}, \bibinfo {author} {\bibfnamefont {F.}~\bibnamefont
  {Pientka}}, \bibinfo {author} {\bibfnamefont {R.}~\bibnamefont {Schmidt}},
  \bibinfo {author} {\bibfnamefont {G.}~\bibnamefont {Zarand}}, \bibinfo
  {author} {\bibfnamefont {E.}~\bibnamefont {Demler}},\ and\ \bibinfo {author}
  {\bibfnamefont {A.}~\bibnamefont {Imamoglu}},\ }\bibfield  {title} {\bibinfo
  {title} {Transport of neutral optical excitations using electric fields},\
  }\href@noop {} {\bibfield  {journal} {\bibinfo  {journal} {Physical Review
  X}\ }\textbf {\bibinfo {volume} {9}},\ \bibinfo {pages} {041019} (\bibinfo
  {year} {2019})}\BibitemShut {NoStop}%
\bibitem [{\citenamefont {Kadanoff}(2018)}]{kadanoff2018quantum}%
  \BibitemOpen
  \bibfield  {author} {\bibinfo {author} {\bibfnamefont {L.~P.}\ \bibnamefont
  {Kadanoff}},\ }\href@noop {} {\emph {\bibinfo {title} {Quantum statistical
  mechanics}}}\ (\bibinfo  {publisher} {CRC Press},\ \bibinfo {year}
  {2018})\BibitemShut {NoStop}%
\bibitem [{\citenamefont {Landau}\ and\ \citenamefont
  {Lifshitz}(2013)}]{landau2013quantum}%
  \BibitemOpen
  \bibfield  {author} {\bibinfo {author} {\bibfnamefont {L.~D.}\ \bibnamefont
  {Landau}}\ and\ \bibinfo {author} {\bibfnamefont {E.~M.}\ \bibnamefont
  {Lifshitz}},\ }\href@noop {} {\emph {\bibinfo {title} {Quantum mechanics:
  non-relativistic theory}}},\ Vol.~\bibinfo {volume} {3}\ (\bibinfo
  {publisher} {Elsevier},\ \bibinfo {year} {2013})\BibitemShut {NoStop}%
\bibitem [{\citenamefont {Levitov}\ and\ \citenamefont
  {Shytov}(2003)}]{levitov2003green}%
  \BibitemOpen
  \bibfield  {author} {\bibinfo {author} {\bibfnamefont {L.~S.}\ \bibnamefont
  {Levitov}}\ and\ \bibinfo {author} {\bibfnamefont {A.~V.}\ \bibnamefont
  {Shytov}},\ }\bibfield  {title} {\bibinfo {title} {{Green's functions.
  Theory and practice}},\ }\href@noop {} {\bibfield  {journal} {\bibinfo
  {journal} {Russian, \href{http://www.mit.edu/~levitov/book}{http://www.mit.edu/\textasciitilde levitov/book}}\ } (\bibinfo {year}
  {2003})}\BibitemShut {NoStop}%
\end{thebibliography}

\end{document}